\documentclass[aps,prd,preprint,preprintnumbers,showpacs,showkeys,superscriptaddress,nofootinbib,floatfix]{revtex4-1}
\usepackage{graphicx}
\usepackage{epsfig,amssymb,amsmath,color}
\usepackage{epstopdf}
\usepackage[colorlinks=true,citecolor=blue,linkcolor=blue,dvipdfmx]{hyperref}

\newcommand{\al}[1]{\begin{align}#1\end{align}}
\newcommand{\bea}{\begin{eqnarray}}   
\newcommand{\eea}{\end{eqnarray}}
\newcommand{\bear}{\begin{array}}  
\newcommand {\eear}{\end{array}}
\newcommand{\bef}{\begin{figure}}  
\newcommand {\eef}{\end{figure}}
\newcommand{\bec}{\begin{center}}  
\newcommand {\eec}{\end{center}}

\def\GEV#1{10^{#1}{\rm\,GeV}}

\newcommand{\ns}{N_{\rm source}}
\newcommand{\na}{N_{\rm axion}}

\begin{document}
\title{Axion Landscape and Natural Inflation}

\author{Tetsutaro Higaki}
\email{thigaki@post.kek.jp}
\affiliation{Theory Center, KEK, 1-1 Oho, Tsukuba, Ibaraki 305-0801, Japan}

\author{Fuminobu Takahashi}
\email{fumi@tuhep.phys.tohoku.ac.jp}
\affiliation{Department of Physics, Tohoku University, Sendai 980-8578, Japan}
\affiliation{Kavli Institute for the Physics and Mathematics of the
  Universe (WPI), Todai Institutes for Advanced Study, University of Tokyo,
  Kashiwa 277-8583, Japan}

\begin{abstract}
Multiple axions form a landscape in the presence of various shift symmetry breaking terms.
Eternal inflation populates the axion landscape, continuously creating new universes by bubble nucleation. 
Slow-roll inflation takes place after the tunneling event, if a very flat direction with 
a super-Planckian  decay constant arises due to the alignment mechanism.
We study the vacuum structure as well as possible inflationary dynamics in the axion landscape scenario, and find that
the  inflaton dynamics is given by either natural or multi-natural inflation. In the limit of large
decay constant, it is approximated by the quadratic chaotic inflation, which however is disfavored if
there is a pressure toward shorter duration of inflation. Therefore,
if the spectral index and the tensor-to-scalar ratio turn out to be different
from the quadratic chaotic inflation, there might be observable traces of the bubble nucleation. 
Also, the existence of small modulations to the inflaton potential is a 
common feature  in the axion landscape, which generates a sizable and almost constant 
running of the scalar spectral index over CMB scales. 
Non-Gaussianity of equilateral type can also be
generated if some of the axions are coupled to massless gauge fields.

\end{abstract}
\preprint{KEK-TH-1772, IPMU14-0308, TU-983}
\maketitle

\section{Introduction}
Our Universe probably experienced the inflationary expansion at an early stage of the 
evolution~\cite{Guth:1980zm,Sato:1980yn,Starobinsky:1980te,Brout:1977ix,Kazanas:1980tx}.
The temperature and polarization anisotropies of the cosmic microwave background (CMB) 
encode the detailed information of the slow-roll inflationary dynamics~\cite{Linde:1981mu,Albrecht:1982wi}.
In particular, if the primordial
B-mode polarization of the CMB is detected, it will be an important milestone toward a proof of inflation.

Recently the BICEP2 collaboration announced the detection of the B-mode polarization
which could be due to the primordial gravitational wave with tensor-to-scalar ratio 
$r = 0.20^{+0.07}_{-0.05}$~\cite{Ade:2014xna}. 
The level of  dust contamination, however, has turned out to be rather high 
according to the {\it Planck} polarization data at $353$\,GHz~\cite{Adam:2014oea}, and 
an ongoing joint analysis of the {\it Planck} and BICEP2 data sets will clarify how much
of the BICEP2 signal is due to the polarized dust emission. Still,
a small but non-negligible fraction of the detected B-mode polarization might be due to the 
tensor mode; for instance, $r \sim 0.1$ may be allowed or 
even preferred~\cite{Colley:2014nna} after subtraction of the dust contribution. Furthermore, even if the large fraction
of the BICEP2 signal arises from dust emission,  $r = 0.01 \sim 0.1$ will still be allowed~\cite{Mortonson:2014bja,Cheng:2014pxa}, and 
$r$ of this order would point definitively 
to large field inflation such as chaotic inflation~\cite{Linde:1983gd},  where the inflaton field 
excursion exceeds the  Planck scale~\cite{Lyth:1996im}. Thus, there is still a room for large-field
inflation, which is consistent with the current observations and will be probed by future 
satellite, balloon-borne, and ground-based CMB polarization experiments.

The central issue in large field inflation models is how to control the inflaton potential over super-Planckian 
field ranges. One plausible possibility is to introduce an approximate shift symmetry on the inflaton. 
The simplest inflation model along this line is natural inflation with the potential~\cite{Freese:1990rb}
\al{
\label{naturalV}
V(\phi) = \Lambda^4 \left(1-\cos\left(\frac{\phi}{f}\right)\right),}
where the cosine potential is induced by some non-perturbative effects. In order to be consistent with the 
Planck data, the decay constant $f$ must satisfy $f \gtrsim 5 M_P$~\cite{Ade:2013uln}, where $M_P 
\simeq 2.4 \times \GEV{18}$ is the reduced Planck mass.\footnote{
There is no observational lower bound on the decay constant in the multi-natural inflation~\cite{Czerny:2014wza,Czerny:2014xja}.
} The inflationary energy scale is about $2 \times \GEV{16}$ close to the GUT scale.
Thus, it is important to build a concrete inflation model in a UV theory such as string theory (see e.g. Refs.~\cite{Dimopoulos:2005ac,Silverstein:2008sg,
McAllister:2008hb,Kaloper:2008fb,Ashoorioon:2009wa,Kaloper:2011jz,Palti:2014kza,Marchesano:2014mla,Hebecker:2014eua,Arends:2014qca,McAllister:2014mpa},
and also Ref.~\cite{Baumann:2014nda} for a recent view.).

The string theory is a promising candidate of unified theory for describing the quantum gravity.
There appear many axions through compactification of the extra dimensions, 
and some of them may remain relatively light and play an important role in cosmology.
In particular, one of such string axions could be the inflaton with the above potential (\ref{naturalV}).
The required large decay constant, however, is not  straightforward to realize because
the fundamental decay constant of the string axions is no larger than the Planck scale
in the limit of a weak coupling or a large extra dimension
(see e.g. Refs.~\cite{Banks:2003sx,Svrcek:2006yi,Rudelius:2014wla}, also Ref.~\cite{Grimm:2014vva} for a strong coupling case, and Ref.~\cite{Kenton:2014gma} for a case with a warped extra dimension). 

If there are two (or more) axions, the effective decay constant can be enhanced by the so called Kim-Nilles-Peloso (KNP) 
alignment mechanism~\cite{Kim:2004rp}; the effective decay constant can be super-Planckian, even if the original
ones are sub-Planckian. 
The KNP mechanism has attracted much attention
especially after the BICEP2 result and it has been studied from various 
aspects~\cite{Czerny:2014xja,Czerny:2014qqa,Harigaya:2014eta,Choi:2014rja,Higaki:2014pja,
Tye:2014tja,Ben-Dayan:2014zsa,Kappl:2014lra,Long:2014dta,Ben-Dayan:2014lca,Westphal:2014ana,Gao:2014uha}.
The original KNP mechanism~\cite{Kim:2004rp} relied upon two axions with sub-Planckian decay constants,
and a relatively large hierarchy in the anomaly coefficients
was required for successful inflation.
It was shown in Ref.~\cite{Choi:2014rja} that, if there are more than two 
axions, large enhancement is possible even with the anomaly coefficients of order unity. 
The probability for the enhancement was  studied  in a case with hierarchical 
coefficients of the cosine functions~\cite{Choi:2014rja} as well as general cases including the case
in which the number of cosine functions $N_{\rm source}$ is different from the number of axions 
$N_{\rm axion}$~\cite{Higaki:2014pja}.\footnote{
In the most of the literature, the relation $N_{\rm axion} = N_{\rm source}$ was assumed.
}

Multiple axions  form a landscape if there are various shift symmetry breaking terms 
with $\ns > \na$, as the present authors proposed in Ref.~\cite{Higaki:2014pja}. 
Eternal inflation will populate a large number of local minima, continuously creating new universes by bubble nucleation~\cite{Coleman:1980aw}.  
Slow-roll inflation takes place after the tunneling event, if a very flat direction 
with a super-Planckian decay constant arises due to the KNP mechanism. 
Thus, eternal inflation as well as the slow-roll inflation that follows the bubble nucleation can be
realized in a unified manner in the axion landscape, while there is no clear 
connection between these two in the string landscape paradigm~\cite{Bousso:2000xa, Susskind:2003kw}.
Furthermore, since the flat direction appears accidentally, there is a pressure toward shorter duration of the slow-roll 
inflation~\cite{Higaki:2014pja}.\footnote{
The pressure arises from the fact that flatter directions (i.e. larger decay constants) are rarer in the KNP mechanism 
with multiple axions. Note however that the existence of such pressure actually depends on the cosmological measures.
} If the total
e-folding number is just about $50$ or $60$, it may be possible to observe some traces of 
the bubble nucleation such as negative curvature~\cite{Linde:1995xm,Linde:1995rv,Freivogel:2005vv,Sugimura:2011tk,Freivogel:2014hca} 
and/or suppression of density perturbations at large scales~\cite{Yamauchi:2011qq,Bousso:2013uia,Bousso:2014jca,Murayama:2014saa}.

The purpose of this paper is to study further the vacuum structure as well as possible inflationary dynamics in the axion landscape,
both of which have not been examined in detail so far. 
We will first show that there are indeed numerous local minima  in the axion landscape and that the energy 
density at local minima approaches a Gaussian distribution as $\ns$ becomes 
larger for a fixed $\na$. Next we examine the inflaton potential along the lightest direction in the axion landscape.
As we shall see below, the possible inflaton dynamics depends on the properties of the cosine functions
and the strength of pressure toward shorter duration of inflation. As a result,  the predicted values of the 
spectral index and the tensor-to-scalar
ratio are correlated with the existence or absence of the measurable remnant of the bubble nucleation.

\section{Axion Landscape}
We consider multiple axions with the following potential:
\al{
\label{axionV}
V(\phi_\alpha) = \sum_{i=1}^{\ns} \Lambda_i^4  
\left(1-\cos\left(\sum_{\alpha=1}^{\na} n_{i\alpha} \frac{\phi_\alpha}{f_\alpha} + \theta_i \right)\right) + C,
}
where  $\Lambda_i$ represents the dynamical scale of each non-perturbative effect,
$n_{i \alpha}$ is an integer-valued anomaly coefficient matrix, $f_\alpha$ is the decay constant,  $\phi_\alpha$ is the 
axion which satisfies
a discrete shift symmetry,
\al{
\phi_\alpha \rightarrow \phi_\alpha + 2 \pi f_\alpha,
}
and $\theta_i \in [0, 2\pi)$ denotes a CP phase of each non-perturbative effect. 
Here the decay constant $f_\alpha$ is
defined so that the greatest common divisor of $\{n_{i\alpha}\}$ is equal to unity for a given $\alpha$. 
The constant term $C$ is chosen so that the cosmological constant almost vanishes in the present Universe.\footnote{
We may need a string landscape to solve the cosmological constant problem unless there is a sufficiently large number
of axions. Then, the axion landscape might be thought of as a lower-energy branch of the string landscape. 
}

Now we study  the vacuum structure in the axion landscape. 
The vacuum structure crucially depends  on the values of $\ns$ and $\na$.
In the case of $\ns < \na$, there are $\na-\ns$ flat directions implying continuous vacuum degeneracy,
as long as the combination of the axions appearing in each cosine function is 
different from one another; if two of the combination are proportional to each other, i.e., $n_{i\alpha}/n_{j \alpha}$
is independent of $\alpha$ for some $i$ and $j$,  there will be another flat direction. 
If $\ns = \na$, there are discrete potential minima with degenerate energy. If $\ns > \na$, there are many
local minima with different energy.  In the following we focus on the last case to examine the distribution of the 
the local minima for various $\ns$ and $\na$.

In order to study the energy density distribution of the local minima, 
we simplify the axion potential (\ref{axionV})
by setting $\Lambda_i = \Lambda$ and $f_\alpha = f$ for all $i$ and $\alpha$ and $C=0$, 
and generate an integer-valued $\ns \times \na$ random matrix $n_{i \alpha}$ satisfying a 
constraint $|n_{i \alpha}| \leq 3$. Note that the structure of the axion potential
does not depend on $C$.
 We have also generated a random real number between
$0$ and $2\pi$ for  the relative phase $\theta_i$. 
Those simplifications are assumed in the following numerical analysis unless stated otherwise.

{ We note that the above-stated simplifications do not change our results significantly, because of the following reasons.
First, cosine functions with $\Lambda_i$ much smaller than the others are irrelevant for the existence
of a very flat direction suitable for the inflaton. Such subdominant cosine functions will generically induce
small modulations to the inflaton potential, which can generate a sizable running of the spectral index,
as we shall see later in this paper. Secondly, the precise equality of $f_\alpha$ is not essential in our analysis, as
we generate random numbers for the coefficients $n_{i \alpha}$, which effectively randomize the size
of the decay constants. Indeed, our analysis on the KNP mechanism can be straightforwardly applied to more general
$\{f_\alpha\}$, which are comparable to one another. This can be seen by noting that we can define
$\theta_\alpha \equiv \phi_\alpha/f_\alpha$ absorbing the decay constants. The precise values of $f_\alpha$
matter only when we evaluate the shape of the inflaton potential quantitatively. 
}

The KNP mechanism can be implemented if there appears an extremely light direction in  the
axion landscape.\footnote{Note that this is because  the typical potential height is of order $\Lambda^4$. If the dynamical scales are
hierarchical,  the lightest direction must be much lighter than the {\it typical} mass scale for the lightest direction,  in order to
realize the KNP mechanism. }
If all the relative phases are zero (or if they can be absorbed by the shift of the axions), this is the case if
the smallest eigenvalues of the matrix ${\cal M}_{\alpha \beta} \equiv \sum_i n_{i \alpha} n_{i \beta}$
happens to be much smaller than the others~\cite{Higaki:2014pja}. Let us denote by $R (\ll 1)$ the ratio of the smallest eigenvalue to 
the next smallest one. Note here that the eigenvalues of ${\cal M}$ are non-negative.
In general, all the relative phases cannot be absorbed by the shift of the axions if $\ns > \na$. Still,
as we shall see shortly,  when $R \ll1$, there appears a very light direction whose typical decay constant
is enhanced by a factor of $1/\sqrt{R}$.
This can be understood as follows. First let us consider the case of $\ns=\na$. Then, ${\cal M}_{\alpha \beta}$ is proportional to
the mass matrix at the points where all the cosine functions are minimized. Since the typical potential height is of order $\Lambda^4$,
 $R \ll 1$ implies that the lightest direction has an enhanced decay constant. Now let us add an extra cosine function of some combination of the axions.
Then, the matrix ${\cal M}_{\alpha \beta}$ is no longer proportional to the mass matrix in the actual axion landscape, because all the 
cosine functions can not be generically minimized simultaneously.  In this case, $R \ll 1$ implies that the combination of the axions 
appearing in the extra cosine function should be more or less orthogonal to the lightest direction obtained in the case of $\ns=\na$.
Thus, the KNP mechanism can be implemented by requiring $R \ll1 $ even for the case of $\ns >\na$ with a non-zero relative phase. 
This argument will be valid unless $\ns-\na$ is not significantly larger than $\na$. 
In the following we use this condition to generate an axion landscape where there is a flat direction, but
our results do not depend on how we implement the KNP mechanism in the axion landscape. 

\begin{figure}[t!]
\begin{center}
\includegraphics[width=7cm]{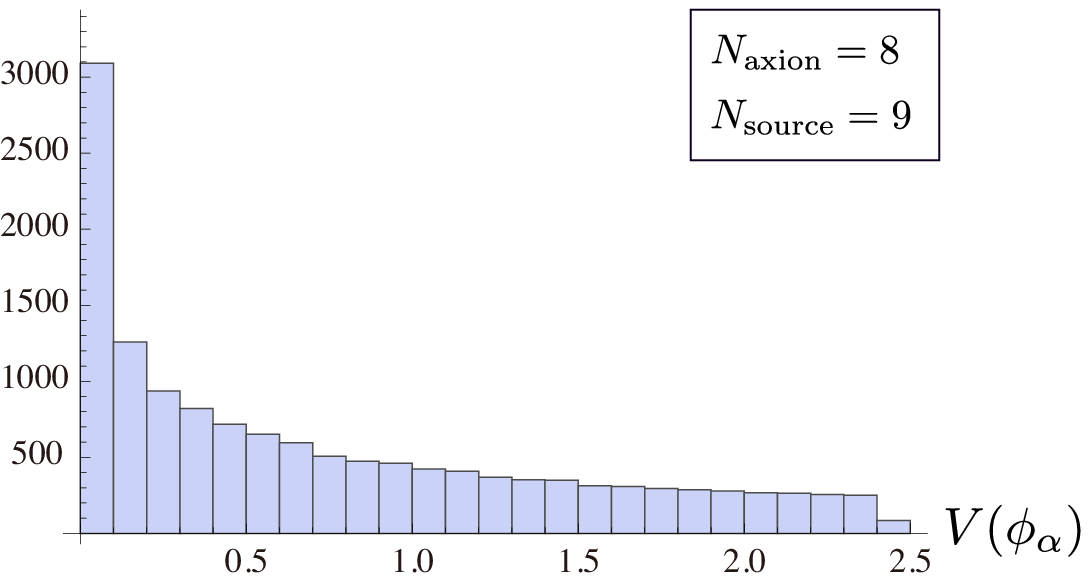}
\includegraphics[width=7cm]{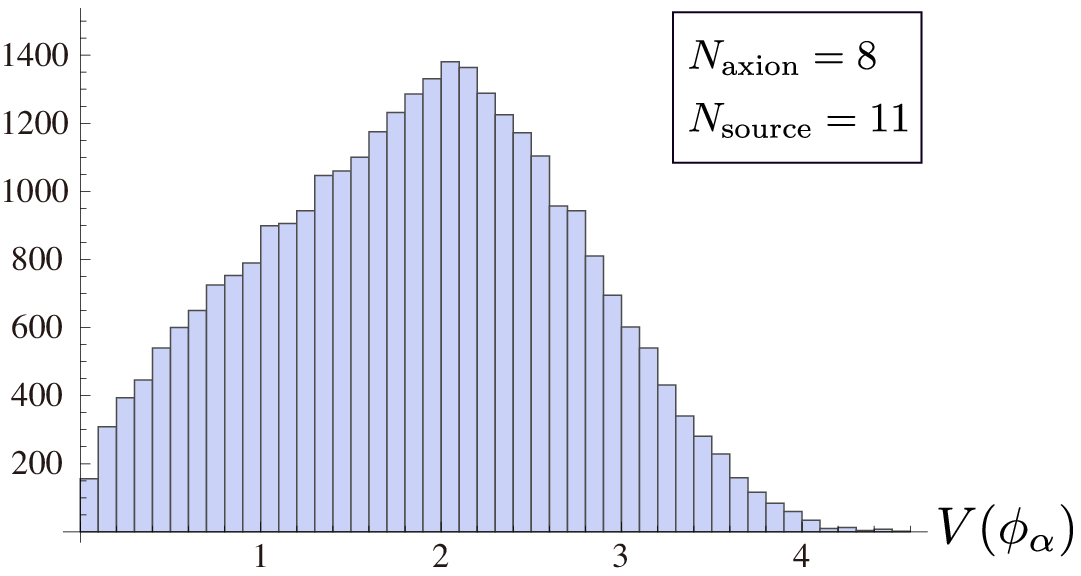}
\includegraphics[width=7cm]{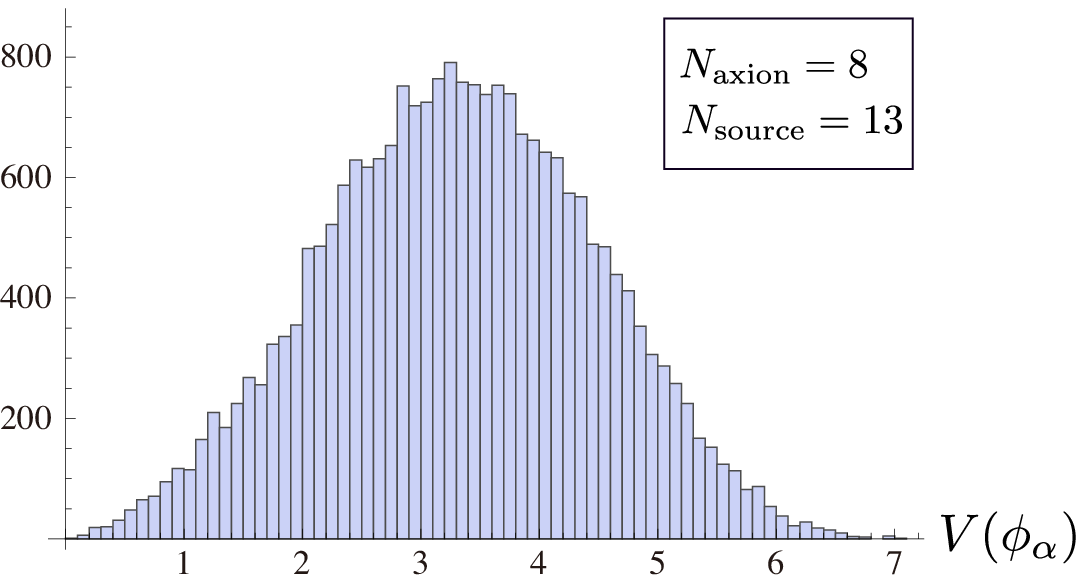}
\caption{
Distribution of axion potential $V(\phi_\alpha)$ at the found local minima in the units of $\Lambda^4$. 
We set $\na = 8$, and we take $\ns = 9, 11$ and $13$ from left to right. Each distribution is for
one realization of the axion landscape satisfying $R < 10^{-2}$.
}
\label{energydistribution}
\end{center}
\end{figure}

The energy density distribution of the local minima is shown in Fig.~\ref{energydistribution}.
Here the number of axions is fixed to be $\na = 8$, and we have varied the number of cosine functions 
as $\ns = 9, 11$ and $13$.  We have repeatedly generated the random matrix for $n_{i \alpha}$ until
$R$ becomes smaller than $10^{-2}$ so as to implement the KNP mechanism.
In each case we have searched for local minima 
in the vicinity of a randomly chosen initial position in the field space, and we have repeated this process until 
distribution of the found local minima converges.
As a result we have found $14025$, $30194$ and $22939$ minima for $\ns = 9, 11$ and $13$, respectively.
We can see from the figure that
the distribution approaches a Gaussian distribution as $\ns$ increases for a fixed $\na$. 
The number of the local minima tend to increase and the energy density distribution 
approaches a Gaussian distribution as $\na$ and the upper bound on $|n_{i \alpha}|$ increase.
We have confirmed that, even if we do not impose the  KNP mechanism,  the energy density distribution of the local minima exhibits 
a similar behavior.

Our vacuum may or may not be around the peak of the energy distribution. This crucially depends on the prior probability distribution
of the parameters in the axion landscape as well as  the constant $C$ which represents the other contributions to the cosmological
constant. This issue is also related to the initial condition of the slow-roll inflation, but we do not pursue it further here.

\begin{figure}[t!]
\begin{center}
\includegraphics[width=7cm]{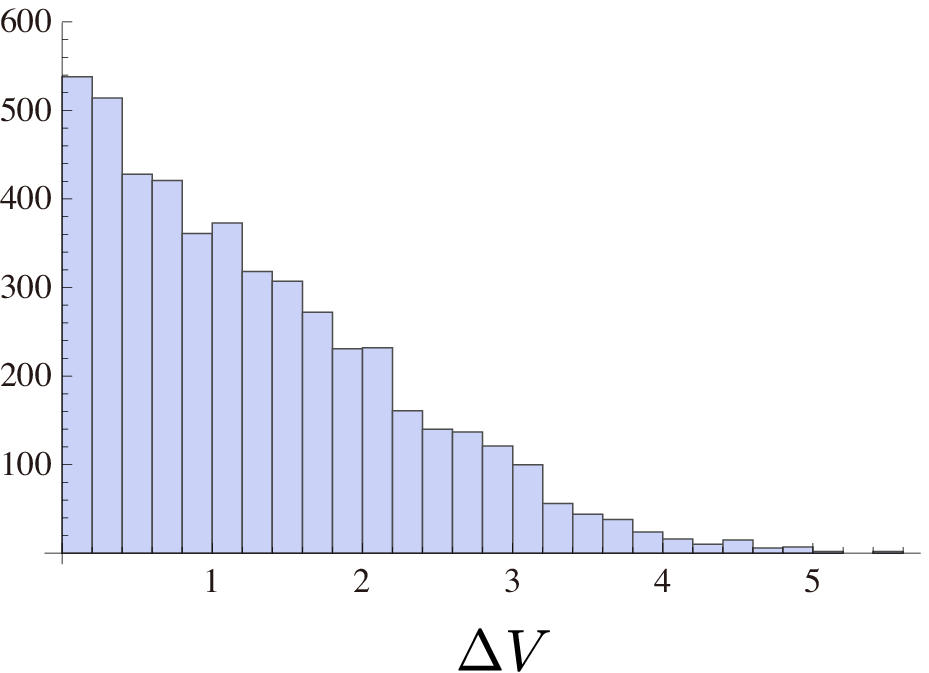}
\includegraphics[width=7cm]{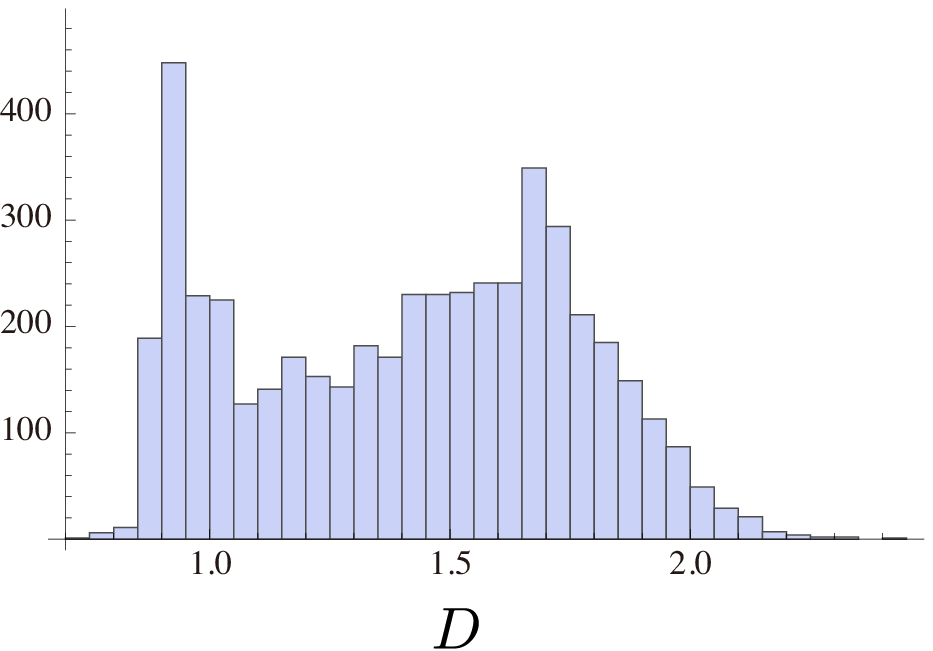}
\caption{
Distribution of the difference of energy density ($\Delta V$) and the distance $D$ 
between each local minimum and its nearest adjacent one, in the units of $\Lambda^4$ and $f$, respectively. 
Here we used the axion landscape with $\na = 8$ and $\ns = 13$ in Fig.~\ref{energydistribution}.
The mean energy difference and distance are about $1.3 \Lambda^4$ and $1.4 f$, respectively. 
}
\label{energydif}
\end{center}
\end{figure}

The eternal inflation takes place if the Universe is stuck in one of the local minima with a positive energy. 
The bubble nucleation rate crucially depends on the energy difference between the adjacent vacua. We have evaluated the
difference of the energy density and the distance between each local minimum and its nearest adjacent one in the case of
$\na=8$ and $\ns=13$ studied above.  The results are shown in Fig.~\ref{energydif}.
One can see that, while there is a peak at the vanishing energy density difference, the typical energy difference
is of order unity in the units of $\Lambda^4$.   
Similarly, the typical distance between the two adjacent minima is of order $f$. The non-trivial distribution of the distance  reflects the structure
of the axion landscape in this example, and it is not due to artifact of the imposed KNP mechanism.\footnote{We have confirmed that the detailed 
distance distribution depends on the realization of the axion landscape, but it is always the case that the typical distance between
adjacent vacua is given by ${\cal O}(f)$. }
 The averaged energy density difference and distance are about $1.3 \Lambda^4$ 
and $1.4 f$, respectively.  This is important for the initial condition of the subsequent slow-roll inflation, and we shall 
return to this issue later.

\section{Inflation in axion landscape}
A very light axion with an effective super-Planckian decay constant can be realized with a certain probability in the presence 
of multiple axions owing to the KNP mechanism~\cite{Choi:2014rja,Higaki:2014pja}. 
Now we examine the inflaton potential in the axion landscape, where there are various
shift-symmetry breaking terms with $\ns > \na$.

\begin{figure}[t!]
\begin{center}
\includegraphics[width=10cm]{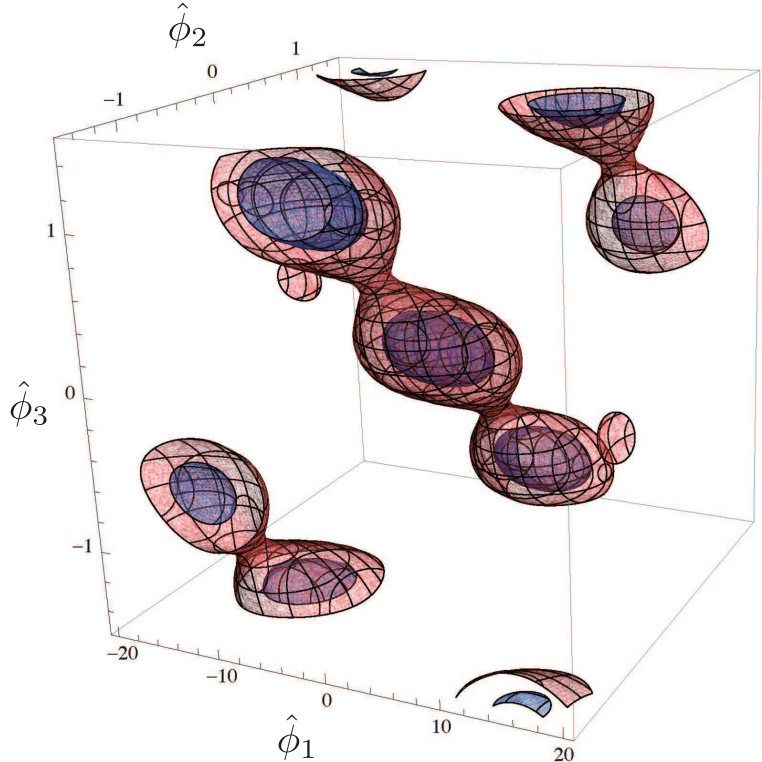}
\caption{
The three dimensional contour plot of the axion potential when the KNP mechanism is operative. The three axis $(\hat{\phi}_1, \hat{\phi}_2, \hat{\phi}_3)$ 
are  chosen so that they coincide with the mass eigenstates at one of the local minima at the origin.  
Note that the scale of each axis is different, and the flat direction is approximately along the direction of $\hat{\phi}_1$.
}
\label{contour}
\end{center}
\end{figure}

First let us show how the axion potential looks like when the KNP mechanism is operative. 
For visualization purpose we set $\na = 3$ and $\ns = 4$ and repeatedly generated an integer-valued matrix 
$n_{i \alpha}$ until $R$ becomes smaller than $10^{-2}$.
 In Fig.~\ref{contour} the surfaces show particular values of the axion potential, 
 $V(\hat{\phi})=2 \Lambda^4$ (purple) and $V(\hat{\phi}) = 2.5 \Lambda^4$ (orange),
 where the three axis $(\hat{\phi}_1, \hat{\phi}_2, \hat{\phi}_3)$
are chosen so that they coincide with the mass eigenstates at one of the local minima, which is
shifted to coincide with the origin ($\hat{\phi}_1=\hat{\phi}_2=\hat{\phi}_3=0$) in the figure.  
Each shaded (purple) ellipsoid  contains one local minimum, while the light shaded (orange)
casing represents the flat direction  in the field space. 
Note that the scale of each axis is different; the range of each axis shown in this figure is $|\hat{\phi}_1|<20 f$ and $|\hat{\phi}_{2,3}| < 1.5 f$.
Therefore the potential is 
indeed flat (approximately) along the lightest direction ($\hat{\phi}_1$), while the other directions ($\hat{\phi}_{2,3}$) are heavier. 

 The inflaton is identified with the lightest degrees of freedom, while the other heavier degrees of freedom
can be integrated out during inflation. There are many local minima and a light direction is 
attached to each minimum. 
As $\na$ increases, it becomes easier to obtain such a flat direction~\cite{Higaki:2014pja} when the relevant dynamical scales are 
comparable to each other.
Thus, the large-field  slow-roll inflation after the bubble nucleation is a  possible outcome in the axion landscape. 

A couple of comments are in order. First the mass eigenstates as well as the mass eigenvalues generically
depend on the position in the field space. Therefore, as one moves away from the local minima along 
the lightest direction,  the composition of the inflaton gradually changes, and as a consequence, the inflaton potential is slightly 
modified from the simple cosine function. Rather, it is given by a certain superposition of the multiple
cosine functions, i.e., multi-natural inflation~\cite{Czerny:2014wza}.  This effect becomes significant especially if $\ns > \na$, because each local minimum
has a different energy, in general. However, if the enhancement of the effective decay constant 
due to the KNP mechanism is large
enough, the inflaton potential can be approximately given by the natural inflation during the last $50$ or $60$ e-foldings, and in the limit of
large enhancement it is approximated by the quadratic chaotic inflation. On the other hand, 
an extremely large decay constant is rare in the axion landscape.\footnote{That said, it is hard to make a definite statement
concerning the probability because of the measure problem.} Thus, it depends on the pressure toward shorter
inflation in the landscape whether the effective inflaton potential is given by the quadratic chaotic inflation or
natural inflation. If the pressure is strong enough and if it persists even after taking account of the cosmological measures, 
the total duration of the inflation is likely close to just about $50$ to $60$,
and we expect that both deviation from the quadratic chaotic inflation and the traces of the bubble nucleation
may be observed together.\footnote{
We assume an anthropic sharp lower bound on the e-folding number around $N_e = 50-60$. 
} The combination of these two observations is one possible outcome of the axion landscape. 

\begin{figure}[t!]
\begin{center}
\includegraphics[width=12cm]{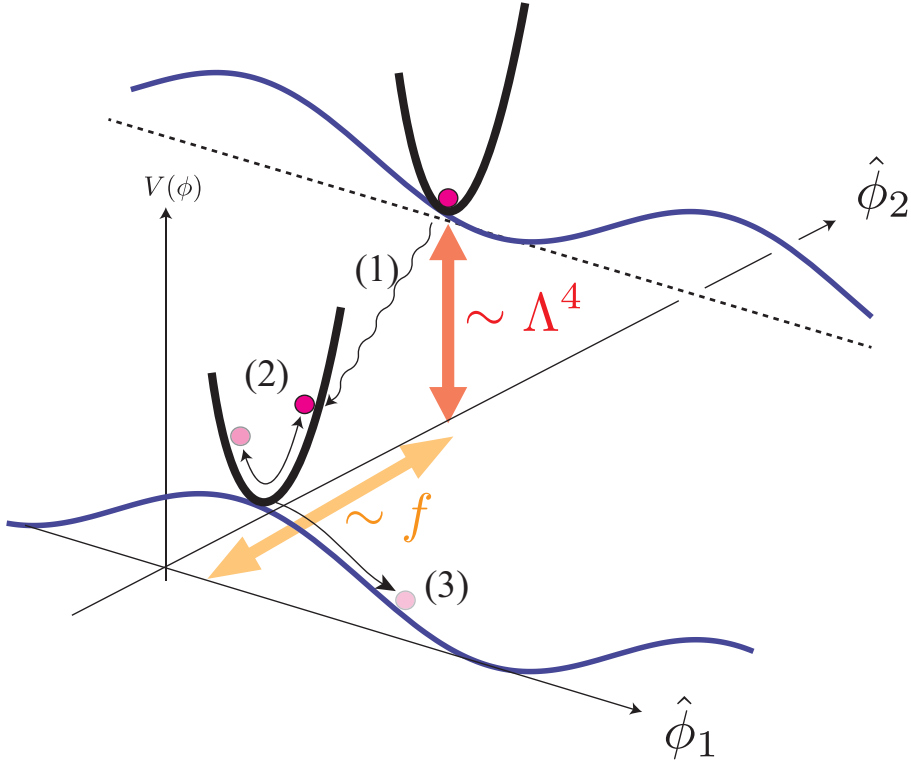}
\caption{
The schematic view of eternal inflation and subsequent slow-roll inflation in the axion landscape. The flat direction $\hat{\phi}_1$ arises
due to the KNP mechanism and $\hat{\phi}_2$ collectively represents the other heavy modes.  First, the Universe is stuck in one of the
local minima, and eternal inflation takes place. 
 After a certain point of time,  (1) tunneling takes place, (2)  heavy axions fast roll and oscillate, and (3) slow-roll inflation starts
along the flat direction.
}
\label{inflation}
\end{center}
\end{figure}

Secondly, the  position of the inflaton after the tunneling event 
is expected to be away from the local minimum by a factor of the effective decay constant along the flat direction. 
This is because the lightest direction does not participate the tunneling event, and there is no special reason for the inflaton 
to just sit on the local minimum after the tunneling~\cite{Linde:1995xm}. One can also understand this from the structure of the axion landscape. 
Suppose that there is a very flat direction owing to the KNP mechanism, and that the Universe experiences eternal inflation
when it is stuck in one of the local minima. Then, after a certain point of time, a bubble forms and the Universe tunnels to some point
along the valley with an energy lower by ${\cal O}(\Lambda^4)$. Therefore, from the energy conservation point of view, 
the axion fields at the center of the bubble can be anywhere along
the valley.  See Fig.~\ref{inflation} for the schematic picture of the eternal inflation and subsequent slow-roll inflation in the axion landscape.
On the other hand, if the energy difference between the two local minima were much smaller than $\Lambda^4$, the tunneling point would be
limited to the vicinity of the lower local minimum. 
Thus, the typical duration of the slow-roll inflation after the bubble formation is determined by the effective decay constant
along the lightest direction. If the effective decay constant is of order $f=5 \sim 10 M_P$, the total duration of inflation could be
just $50$ or $60$.

We have so far assigned a common value $\Lambda$ to  all the dynamical scales. However, there is no special reason
to expect that this is the case, and indeed, some of the dynamical scales, $\Lambda_L$,  can be (much) smaller than the
others,  $\Lambda_L \lesssim \Lambda$.  Then, such cosine functions do not participate in the KNP mechanism, but they
would give rise to small modulations to the
inflaton potential. To be concrete, let us suppose that all the decay constants are of similar order, say, $f_\alpha = f 
\approx \GEV{17}$, and the effective decay constant for the inflaton,  $f_{\rm eff}$,   is enhanced by a factor of
about $10^2$ owing to the KNP mechanism, i.e., $f_{\rm eff} = {\cal O}(10^2) f = {\cal O}(10)M_P$. The small 
modulations are expected to have a decay constant of order $f$, as they do not participate in the KNP mechanism. 

\begin{figure}[t!]
\begin{center}
\includegraphics[width=8cm]{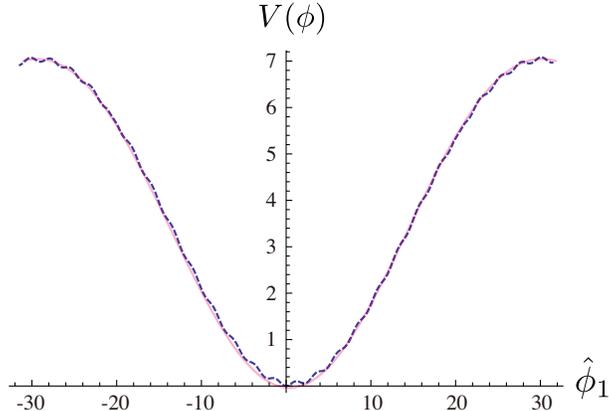}
\caption{
The potential along the lightest axion direction in the axion landscape with (solid)  or without (dashed) small modulations,
which is due to an extra cosine function with a suppressed amplitude. 
}
\label{inflatonpotential}
\end{center}
\end{figure}

In Fig.~\ref{inflatonpotential} we show the axion potential along the lightest axion in the landscape
with $\na = 3$,  $\ns = 5$, $\Lambda_i =  \Lambda$ for $i=1 - 4$ and $\Lambda_5 = 0.1 \Lambda$.
We required $R < 1/20$ to implement the KNP mechanism when the fifth cosine function is absent. 
One can see that there appear small modulations due to the mild hierarchy in the dynamical scales.
Interestingly, such small modulations with  mild hierarchy in the decay constants leads to a sizable running of the scalar 
spectral index,
  $|dn_s/d \ln k |= {\cal O}(0.01)$, which is almost constant
over the CMB scales, and so, there is no contradiction with large-scale 
observations~\cite{Kobayashi:2010pz,Takahashi:2013tj,Czerny:2014wua,Abazajian:2014tqa,Garrison-Kimmel:2014kia}. 
The presence of such small modulations, and therefore the running spectral index, is a common feature
of inflation in the axion landscape. Such periodic modulations to the inflaton potential, if found,  would definitively demonstrate that the inflaton is
one of the axions with softly broken shift symmetry.

\section{Discussion and Conclusions}

After inflation, the inflaton needs to decay and reheat the standard model particles. 
This can be realized if one (or more) of the axions is coupled to the standard model sector via e.g. 
\al{{\cal L} = \frac{\phi_\alpha}{f_\alpha} F_{\mu \nu} \tilde{F}^{\mu \nu}.
\label{phiFFtilde}
}
Here, $F_{\mu \nu}$ is the gauge field strength in the Standard Model.
The decay temperature is estimated to be of order $10^{10}$ GeV with the inflaton mass 
$m_{\phi_{\alpha}} \approx 10^{13}$ GeV and $f_{\alpha} \approx 10^{17}$ GeV.
Thermal leptogenesis will be possible for such a high reheating temperature~\cite{Fukugita:1986hr}.
Note that the decay process should be considered in the mass eigenstate basis, and the couplings
may be suppressed by a mixing angle compared to a naive estimate
(such a possibility is studied in the context of the axion-like particle dark matter \cite{Higaki:2014qua}).
Non-thermal leptogenesis may also be
possible through the inflaton decay into the lightest right-handed neutrino $\nu^c$ with the interaction
${\cal L} = c (\phi_{\alpha}/f_{\alpha}) M \nu^c \nu^c $, where $M$ represents the right-handed neutrino mass
and $c$ is a coupling constant.

The couplings of the inflaton like (\ref{phiFFtilde}) are known to induce non-Gaussianity  of equilateral type if the 
prefactors of $F \tilde{F}$ change significantly during inflation~\cite{Barnaby:2010vf,Barnaby:2011vw}.
For the decay constant of order $f_{\alpha} \approx 10^{17}$ GeV, the current non-Gaussian constraint
can be satisfied. If some of the decay constants are smaller, or if the observational bound on the non-Gaussianity 
is improved, it may be possible to detect the non-Gaussianity generated through the interaction (\ref{phiFFtilde}).

So far we have concentrated on the axions with the potential (\ref{axionV}), assuming the other degrees of freedom 
such as their scalar partner ``saxions" are much heavier. In fact, however, the saxions do not have to be hierarchically 
heavier, since the inflaton (the lightest axion) is much lighter 
than the typical axion mass when the KNP mechanism is operative. It is beyond the scope of this paper
to study how light the saxions can be without modifying our results, but we expect that the basic features of the axion
landscape scenario (eternal inflation, subsequent slow-roll inflation due to the KNP mechanism, the inflaton potential, etc.)
will remain valid even if the saxions have a mass comparable to the axions.
 In the string theory, such an issue is related to the decompactification problem \cite{Kallosh:2004yh}.

Here let us comment on how the decay constant of string theoretic axions is determined.
The size of the decay constant  depends not only on the wave function of an axion (tensor field) in the extra dimension, 
but also on how the axion couplings to gauge fields appear in four dimensions.
The typical size of the decay constant is given by the string scale, the Kaluzua-Klein scale, or the
winding scale. Further, the decay constant can be enhanced  by one-loop factor, 
when the axion couples to the gauge field at the quantum level as in the case of
K\"ahler moduli on a Calabi-Yau space in Heterotic string~\cite{Abe:2014pwa} or type IIA string.

Complex structure moduli in type IIB models also can induce the inflation.
It depends on the detailed moduli stabilization with flux compactification whether such axions drive
natural inflation.  In particular, stringy corrections are important, and
direct couplings of axions to branes/fluxes can induce explicit shift-symmetry breaking terms
which cause the so-called monodromy inflation instead~\cite{Silverstein:2008sg,McAllister:2008hb,Kaloper:2008fb,Kaloper:2011jz,
Palti:2014kza,Marchesano:2014mla,Hebecker:2014eua,Arends:2014qca,McAllister:2014mpa,Hayashi:2014aua,Hebecker:2014kva}.
Although we have assumed canonical kinetic terms for the axions,
the axion kinetic terms may depend on the axions through stringy or non-perturbative corrections.
Then, the inflaton potential will be modified when expressed in terms of the canonically normalized axion fields. 
 Further, depending on the origin of non-perturbative effects,
several issues may arise. For instance, the other tunneling events may take place:
domain walls will be generated through the gaugino condensation in supersymmetric theories \cite{Dine:2014hwa,Yonekura:2014oja}, although then
monodromies, which make the axion potential sufficiently flatter, can be observed in large $N$ 
gauge theories \cite{Witten:1980sp,Witten:1998uka}.
In addition, the coupling of axion to many light modes might make decay constant larger through quantum corrections including them \cite{Yonekura:2014oja}.\footnote{ See also \cite{ArkaniHamed:2006dz},
in which $M_P \gtrsim f$ is required.}

The tadpole condition is another important issue for implementing the KNP mechanism in the string theory, 
which requires a number of (brane) charges through the alignment of wrapping numbers of
branes or gauge fluxes on branes in the extra dimensions, 
while the net charge of branes should be vanishing in the compact extra dimension.\footnote{
Natural inflation with an open string state can be similar, because many D-branes \cite{Pajer:2008uy} or an 
interplay between winding of branes and flux on branes \cite{Kenton:2014gma} 
are required to realize a large decay constant. See also \cite{Harigaya:2014eta} for a 
field theoretical approach, in which a large decay constant is realized with a large charge.}
In this respect, it is interesting 
to consider a possibility that the inflaton consists of RR two-from axions in type IIB model
\cite{Ben-Dayan:2014zsa,Long:2014dta,Gao:2014uha,Ben-Dayan:2014lca}
and fluxed seven branes, which produce non-perturbative inflaton potential,
are wrapping on a curved extra dimension with orientifolds. In this case, a tight consistency 
condition on three brane charges can be relaxed by the curvature corrections
due to the seven branes \cite{Denef:2008wq,Collinucci:2008pf}.
Note that fluxes on the seven branes do not contribute to the seven brane charges
and, therefore, it may ease the constraint on five brane charges if the flux contribution to the anomaly 
is canceled between branes and their mirror images~\cite{Blumenhagen:2008zz}. 
Euclidean brane instantons may also relax this constraint 
because they will not produce real brane charges. Similarly, 
it is interesting to realize natural inflation with IIA/Heterotic K\"ahler moduli 
or IIB complex structure moduli through world-sheet instanton or its dual (classical) effect, respectively;
they have another advantage that the decay constant is enhanced by one-loop factor.
 Wrapping branes for the KNP mechanism 
could also lead to stringy light modes, whose effects are discussed in 
Ref.~\cite{ArkaniHamed:2005yv,Dimopoulos:2005ac,Kaloper:2011jz}:
They will correct the slow-roll parameters via
the Planck scale renormalization, and the vacuum energy during the inflation. Note that the Planck scale is an apparent scale in the string theory.

The QCD axion may be a part of the axion landscape,  but the isocurvature perturbations will be generically too large. 
In fact,  too large isocurvature perturbation is a serious problem for any scenarios including light and cosmological stable
axions or axion-like particles, as long as the inflation scale is high. There are several ways to avoid the isocurvature bound.
One possibility is that the shift symmetry is  broken by a large amount during inflation so that axions are sufficiently 
heavy and their fluctuations are suppressed~\cite{Dine:2004cq,Jeong:2013xta,Higaki:2014ooa}.   Another interesting way
is to make use of MSW-like resonant conversions~\cite{Kitajima:2014xla}.

Throughout this paper we have focused on the large-field inflation in the axion landscape. In fact,
it is also possible to implement small-field inflation based on an axion with  multiple cosine functions.
This is because, for a certain choice of the dynamical scales, the curvature of the axion potential around the hilltop can be
vanishingly small, which leads to an axion hilltop inflation~\cite{Czerny:2014wza,Czerny:2014xja}.
The predicted tensor-to-scalar ratio is smaller; $r \lesssim 10^{-3}$ is expected for the sub-Planckian decay constant.

One of the central roles of the axion landscape is to provide a sufficiently flat direction
suitable for large-field inflation based on the KNP alignment mechanism. If the number
of axions are sufficiently large, the vacuum structure becomes so complicated that
it may help to solve the cosmological constant problem with the aid of  the anthropic 
principle \cite{Weinberg:1988cp}.\footnote{
There is an infinite number of degenerate ground states in the case of
an irrational axion~\cite{Banks:1991mb,Kallosh:2014vja}.}

In this paper we have studied the vacuum structure as well as possible inflationary dynamics in the axion landscape
where there are many axions with various shift-symmetry breaking terms. The required flatness of the inflationary path 
is realized by the KNP alignment mechanism. The axion landscape provides us with a unified understanding of the
eternal inflation in one of the local minima and the subsequent slow-roll inflation after the tunneling event. 
If the deviation from the quadratic chaotic inflation is found, it implies that there is pressure toward shorter duration
of the slow-roll inflation, and so, it may be possible to observe remnants of the bubble nucleation such as negative
spatial curvature as well as the suppression of density perturbations at large scales. The presence of small modulations
to the inflaton potential is common in the axion landscape, which leads to a sizable running of the scalar spectral index
of order $|dn_s/d\ln k| = {\cal O}(0.01)$ which is almost constant over CMB scales.

{\it Note added:}\\
After submission of our paper, the joint analysis of BICEP2/Keck-array and Planck data appeared~\cite{Ade:2015tva}, setting
an upper bound $r < 0.12~(95\%{\rm \,confidence})$. The quadratic chaotic inflation is now disfavored, which
may imply a preference for shorter duration of inflation in our scenario.

\acknowledgments
This work was supported by  JSPS Grant-in-Aid for
Young Scientists (B) (No.24740135 [FT] and No. 25800169 [TH]), 
Scientific Research (A) (No.26247042 [TH, FT]), Scientific Research (B) (No.26287039 [FT]), 
 the Grant-in-Aid for Scientific Research on Innovative Areas (No.23104008 [FT]),  and
Inoue Foundation for Science [FT].  This work was also
supported by World Premier International Center Initiative (WPI Program), MEXT, Japan [FT].

\bibliography{naturalinflation}

\begin{thebibliography}{90}%
\makeatletter
\providecommand \@ifxundefined [1]{%
 \@ifx{#1\undefined}
}%
\providecommand \@ifnum [1]{%
 \ifnum #1\expandafter \@firstoftwo
 \else \expandafter \@secondoftwo
 \fi
}%
\providecommand \@ifx [1]{%
 \ifx #1\expandafter \@firstoftwo
 \else \expandafter \@secondoftwo
 \fi
}%
\providecommand \natexlab [1]{#1}%
\providecommand \enquote  [1]{``#1''}%
\providecommand \bibnamefont  [1]{#1}%
\providecommand \bibfnamefont [1]{#1}%
\providecommand \citenamefont [1]{#1}%
\providecommand \href@noop [0]{\@secondoftwo}%
\providecommand \href [0]{\begingroup \@sanitize@url \@href}%
\providecommand \@href[1]{\@@startlink{#1}\@@href}%
\providecommand \@@href[1]{\endgroup#1\@@endlink}%
\providecommand \@sanitize@url [0]{\catcode `\\12\catcode `\$12\catcode
  `\&12\catcode `\#12\catcode `\^12\catcode `\_12\catcode `\%12\relax}%
\providecommand \@@startlink[1]{}%
\providecommand \@@endlink[0]{}%
\providecommand \url  [0]{\begingroup\@sanitize@url \@url }%
\providecommand \@url [1]{\endgroup\@href {#1}{\urlprefix }}%
\providecommand \urlprefix  [0]{URL }%
\providecommand \Eprint [0]{\href }%
\providecommand \doibase [0]{http://dx.doi.org/}%
\providecommand \selectlanguage [0]{\@gobble}%
\providecommand \bibinfo  [0]{\@secondoftwo}%
\providecommand \bibfield  [0]{\@secondoftwo}%
\providecommand \translation [1]{[#1]}%
\providecommand \BibitemOpen [0]{}%
\providecommand \bibitemStop [0]{}%
\providecommand \bibitemNoStop [0]{.\EOS\space}%
\providecommand \EOS [0]{\spacefactor3000\relax}%
\providecommand \BibitemShut  [1]{\csname bibitem#1\endcsname}%
\let\auto@bib@innerbib\@empty
\bibitem [{\citenamefont {Guth}(1981)}]{Guth:1980zm}%
  \BibitemOpen
  \bibfield  {author} {\bibinfo {author} {\bibfnamefont {A.~H.}\ \bibnamefont
  {Guth}},\ }\href {\doibase 10.1103/PhysRevD.23.347} {\bibfield  {journal}
  {\bibinfo  {journal} {Phys.Rev.}\ }\textbf {\bibinfo {volume} {D23}},\
  \bibinfo {pages} {347} (\bibinfo {year} {1981})}\BibitemShut {NoStop}%
\bibitem [{\citenamefont {Sato}(1981)}]{Sato:1980yn}%
  \BibitemOpen
  \bibfield  {author} {\bibinfo {author} {\bibfnamefont {K.}~\bibnamefont
  {Sato}},\ }\href@noop {} {\bibfield  {journal} {\bibinfo  {journal}
  {Mon.Not.Roy.Astron.Soc.}\ }\textbf {\bibinfo {volume} {195}},\ \bibinfo
  {pages} {467} (\bibinfo {year} {1981})}\BibitemShut {NoStop}%
\bibitem [{\citenamefont {Starobinsky}(1980)}]{Starobinsky:1980te}%
  \BibitemOpen
  \bibfield  {author} {\bibinfo {author} {\bibfnamefont {A.~A.}\ \bibnamefont
  {Starobinsky}},\ }\href {\doibase 10.1016/0370-2693(80)90670-X} {\bibfield
  {journal} {\bibinfo  {journal} {Phys.Lett.}\ }\textbf {\bibinfo {volume}
  {B91}},\ \bibinfo {pages} {99} (\bibinfo {year} {1980})}\BibitemShut
  {NoStop}%
\bibitem [{\citenamefont {Brout}\ \emph {et~al.}(1978)\citenamefont {Brout},
  \citenamefont {Englert},\ and\ \citenamefont {Gunzig}}]{Brout:1977ix}%
  \BibitemOpen
  \bibfield  {author} {\bibinfo {author} {\bibfnamefont {R.}~\bibnamefont
  {Brout}}, \bibinfo {author} {\bibfnamefont {F.}~\bibnamefont {Englert}}, \
  and\ \bibinfo {author} {\bibfnamefont {E.}~\bibnamefont {Gunzig}},\ }\href
  {\doibase 10.1016/0003-4916(78)90176-8} {\bibfield  {journal} {\bibinfo
  {journal} {Annals Phys.}\ }\textbf {\bibinfo {volume} {115}},\ \bibinfo
  {pages} {78} (\bibinfo {year} {1978})}\BibitemShut {NoStop}%
\bibitem [{\citenamefont {Kazanas}(1980)}]{Kazanas:1980tx}%
  \BibitemOpen
  \bibfield  {author} {\bibinfo {author} {\bibfnamefont {D.}~\bibnamefont
  {Kazanas}},\ }\href {\doibase 10.1086/183361} {\bibfield  {journal} {\bibinfo
   {journal} {Astrophys.J.}\ }\textbf {\bibinfo {volume} {241}},\ \bibinfo
  {pages} {L59} (\bibinfo {year} {1980})}\BibitemShut {NoStop}%
\bibitem [{\citenamefont {Linde}(1982)}]{Linde:1981mu}%
  \BibitemOpen
  \bibfield  {author} {\bibinfo {author} {\bibfnamefont {A.~D.}\ \bibnamefont
  {Linde}},\ }\href {\doibase 10.1016/0370-2693(82)91219-9} {\bibfield
  {journal} {\bibinfo  {journal} {Phys.Lett.}\ }\textbf {\bibinfo {volume}
  {B108}},\ \bibinfo {pages} {389} (\bibinfo {year} {1982})}\BibitemShut
  {NoStop}%
\bibitem [{\citenamefont {Albrecht}\ and\ \citenamefont
  {Steinhardt}(1982)}]{Albrecht:1982wi}%
  \BibitemOpen
  \bibfield  {author} {\bibinfo {author} {\bibfnamefont {A.}~\bibnamefont
  {Albrecht}}\ and\ \bibinfo {author} {\bibfnamefont {P.~J.}\ \bibnamefont
  {Steinhardt}},\ }\href {\doibase 10.1103/PhysRevLett.48.1220} {\bibfield
  {journal} {\bibinfo  {journal} {Phys.Rev.Lett.}\ }\textbf {\bibinfo {volume}
  {48}},\ \bibinfo {pages} {1220} (\bibinfo {year} {1982})}\BibitemShut
  {NoStop}%
\bibitem [{\citenamefont {Ade}\ \emph {et~al.}(2014)\citenamefont {Ade} \emph
  {et~al.}}]{Ade:2014xna}%
  \BibitemOpen
  \bibfield  {author} {\bibinfo {author} {\bibfnamefont {P.}~\bibnamefont
  {Ade}} \emph {et~al.} (\bibinfo {collaboration} {BICEP2 Collaboration}),\
  }\href {\doibase 10.1103/PhysRevLett.112.241101} {\bibfield  {journal}
  {\bibinfo  {journal} {Phys.Rev.Lett.}\ }\textbf {\bibinfo {volume} {112}},\
  \bibinfo {pages} {241101} (\bibinfo {year} {2014})},\ \Eprint
  {http://arxiv.org/abs/1403.3985} {arXiv:1403.3985 [astro-ph.CO]} \BibitemShut
  {NoStop}%
\bibitem [{\citenamefont {Adam}\ \emph {et~al.}(2014)\citenamefont {Adam} \emph
  {et~al.}}]{Adam:2014oea}%
  \BibitemOpen
  \bibfield  {author} {\bibinfo {author} {\bibfnamefont {R.}~\bibnamefont
  {Adam}} \emph {et~al.} (\bibinfo {collaboration} {Planck Collaboration}),\
  }\href@noop {} {\  (\bibinfo {year} {2014})},\ \Eprint
  {http://arxiv.org/abs/1409.5738} {arXiv:1409.5738 [astro-ph.CO]} \BibitemShut
  {NoStop}%
\bibitem [{\citenamefont {Colley}\ and\ \citenamefont
  {Gott}(2014)}]{Colley:2014nna}%
  \BibitemOpen
  \bibfield  {author} {\bibinfo {author} {\bibfnamefont {W.~N.}\ \bibnamefont
  {Colley}}\ and\ \bibinfo {author} {\bibfnamefont {J.~R.}\ \bibnamefont
  {Gott}},\ }\href@noop {} {\  (\bibinfo {year} {2014})},\ \Eprint
  {http://arxiv.org/abs/1409.4491} {arXiv:1409.4491 [astro-ph.CO]} \BibitemShut
  {NoStop}%
\bibitem [{\citenamefont {Mortonson}\ and\ \citenamefont
  {Seljak}(2014)}]{Mortonson:2014bja}%
  \BibitemOpen
  \bibfield  {author} {\bibinfo {author} {\bibfnamefont {M.~J.}\ \bibnamefont
  {Mortonson}}\ and\ \bibinfo {author} {\bibfnamefont {U.}~\bibnamefont
  {Seljak}},\ }\href@noop {} {\  (\bibinfo {year} {2014})},\ \Eprint
  {http://arxiv.org/abs/1405.5857} {arXiv:1405.5857 [astro-ph.CO]} \BibitemShut
  {NoStop}%
\bibitem [{\citenamefont {Cheng}\ \emph {et~al.}(2014)\citenamefont {Cheng},
  \citenamefont {Huang},\ and\ \citenamefont {Wang}}]{Cheng:2014pxa}%
  \BibitemOpen
  \bibfield  {author} {\bibinfo {author} {\bibfnamefont {C.}~\bibnamefont
  {Cheng}}, \bibinfo {author} {\bibfnamefont {Q.-G.}\ \bibnamefont {Huang}}, \
  and\ \bibinfo {author} {\bibfnamefont {S.}~\bibnamefont {Wang}},\ }\href@noop
  {} {\  (\bibinfo {year} {2014})},\ \Eprint {http://arxiv.org/abs/1409.7025}
  {arXiv:1409.7025 [astro-ph.CO]} \BibitemShut {NoStop}%
\bibitem [{\citenamefont {Linde}(1983)}]{Linde:1983gd}%
  \BibitemOpen
  \bibfield  {author} {\bibinfo {author} {\bibfnamefont {A.~D.}\ \bibnamefont
  {Linde}},\ }\href {\doibase 10.1016/0370-2693(83)90837-7} {\bibfield
  {journal} {\bibinfo  {journal} {Phys.Lett.}\ }\textbf {\bibinfo {volume}
  {B129}},\ \bibinfo {pages} {177} (\bibinfo {year} {1983})}\BibitemShut
  {NoStop}%
\bibitem [{\citenamefont {Lyth}(1997)}]{Lyth:1996im}%
  \BibitemOpen
  \bibfield  {author} {\bibinfo {author} {\bibfnamefont {D.~H.}\ \bibnamefont
  {Lyth}},\ }\href {\doibase 10.1103/PhysRevLett.78.1861} {\bibfield  {journal}
  {\bibinfo  {journal} {Phys.Rev.Lett.}\ }\textbf {\bibinfo {volume} {78}},\
  \bibinfo {pages} {1861} (\bibinfo {year} {1997})},\ \Eprint
  {http://arxiv.org/abs/hep-ph/9606387} {arXiv:hep-ph/9606387 [hep-ph]}
  \BibitemShut {NoStop}%
\bibitem [{\citenamefont {Freese}\ \emph {et~al.}(1990)\citenamefont {Freese},
  \citenamefont {Frieman},\ and\ \citenamefont {Olinto}}]{Freese:1990rb}%
  \BibitemOpen
  \bibfield  {author} {\bibinfo {author} {\bibfnamefont {K.}~\bibnamefont
  {Freese}}, \bibinfo {author} {\bibfnamefont {J.~A.}\ \bibnamefont {Frieman}},
  \ and\ \bibinfo {author} {\bibfnamefont {A.~V.}\ \bibnamefont {Olinto}},\
  }\href {\doibase 10.1103/PhysRevLett.65.3233} {\bibfield  {journal} {\bibinfo
   {journal} {Phys.Rev.Lett.}\ }\textbf {\bibinfo {volume} {65}},\ \bibinfo
  {pages} {3233} (\bibinfo {year} {1990})}\BibitemShut {NoStop}%
\bibitem [{\citenamefont {Ade}\ \emph {et~al.}(2013)\citenamefont {Ade} \emph
  {et~al.}}]{Ade:2013uln}%
  \BibitemOpen
  \bibfield  {author} {\bibinfo {author} {\bibfnamefont {P.}~\bibnamefont
  {Ade}} \emph {et~al.} (\bibinfo {collaboration} {Planck Collaboration}),\
  }\href@noop {} {\  (\bibinfo {year} {2013})},\ \Eprint
  {http://arxiv.org/abs/1303.5082} {arXiv:1303.5082 [astro-ph.CO]} \BibitemShut
  {NoStop}%
\bibitem [{\citenamefont {Czerny}\ and\ \citenamefont
  {Takahashi}(2014)}]{Czerny:2014wza}%
  \BibitemOpen
  \bibfield  {author} {\bibinfo {author} {\bibfnamefont {M.}~\bibnamefont
  {Czerny}}\ and\ \bibinfo {author} {\bibfnamefont {F.}~\bibnamefont
  {Takahashi}},\ }\href {\doibase 10.1016/j.physletb.2014.04.039} {\bibfield
  {journal} {\bibinfo  {journal} {Phys.Lett.}\ }\textbf {\bibinfo {volume}
  {B733}},\ \bibinfo {pages} {241} (\bibinfo {year} {2014})},\ \Eprint
  {http://arxiv.org/abs/1401.5212} {arXiv:1401.5212 [hep-ph]} \BibitemShut
  {NoStop}%
\bibitem [{\citenamefont {Czerny}\ \emph
  {et~al.}(2014{\natexlab{a}})\citenamefont {Czerny}, \citenamefont {Higaki},\
  and\ \citenamefont {Takahashi}}]{Czerny:2014xja}%
  \BibitemOpen
  \bibfield  {author} {\bibinfo {author} {\bibfnamefont {M.}~\bibnamefont
  {Czerny}}, \bibinfo {author} {\bibfnamefont {T.}~\bibnamefont {Higaki}}, \
  and\ \bibinfo {author} {\bibfnamefont {F.}~\bibnamefont {Takahashi}},\ }\href
  {\doibase 10.1007/JHEP05(2014)144} {\bibfield  {journal} {\bibinfo  {journal}
  {JHEP}\ }\textbf {\bibinfo {volume} {1405}},\ \bibinfo {pages} {144}
  (\bibinfo {year} {2014}{\natexlab{a}})},\ \Eprint
  {http://arxiv.org/abs/1403.0410} {arXiv:1403.0410 [hep-ph]} \BibitemShut
  {NoStop}%
\bibitem [{\citenamefont {Dimopoulos}\ \emph {et~al.}(2008)\citenamefont
  {Dimopoulos}, \citenamefont {Kachru}, \citenamefont {McGreevy},\ and\
  \citenamefont {Wacker}}]{Dimopoulos:2005ac}%
  \BibitemOpen
  \bibfield  {author} {\bibinfo {author} {\bibfnamefont {S.}~\bibnamefont
  {Dimopoulos}}, \bibinfo {author} {\bibfnamefont {S.}~\bibnamefont {Kachru}},
  \bibinfo {author} {\bibfnamefont {J.}~\bibnamefont {McGreevy}}, \ and\
  \bibinfo {author} {\bibfnamefont {J.~G.}\ \bibnamefont {Wacker}},\ }\href
  {\doibase 10.1088/1475-7516/2008/08/003} {\bibfield  {journal} {\bibinfo
  {journal} {JCAP}\ }\textbf {\bibinfo {volume} {0808}},\ \bibinfo {pages}
  {003} (\bibinfo {year} {2008})},\ \Eprint
  {http://arxiv.org/abs/hep-th/0507205} {arXiv:hep-th/0507205 [hep-th]}
  \BibitemShut {NoStop}%
\bibitem [{\citenamefont {Silverstein}\ and\ \citenamefont
  {Westphal}(2008)}]{Silverstein:2008sg}%
  \BibitemOpen
  \bibfield  {author} {\bibinfo {author} {\bibfnamefont {E.}~\bibnamefont
  {Silverstein}}\ and\ \bibinfo {author} {\bibfnamefont {A.}~\bibnamefont
  {Westphal}},\ }\href {\doibase 10.1103/PhysRevD.78.106003} {\bibfield
  {journal} {\bibinfo  {journal} {Phys.Rev.}\ }\textbf {\bibinfo {volume}
  {D78}},\ \bibinfo {pages} {106003} (\bibinfo {year} {2008})},\ \Eprint
  {http://arxiv.org/abs/0803.3085} {arXiv:0803.3085 [hep-th]} \BibitemShut
  {NoStop}%
\bibitem [{\citenamefont {McAllister}\ \emph {et~al.}(2010)\citenamefont
  {McAllister}, \citenamefont {Silverstein},\ and\ \citenamefont
  {Westphal}}]{McAllister:2008hb}%
  \BibitemOpen
  \bibfield  {author} {\bibinfo {author} {\bibfnamefont {L.}~\bibnamefont
  {McAllister}}, \bibinfo {author} {\bibfnamefont {E.}~\bibnamefont
  {Silverstein}}, \ and\ \bibinfo {author} {\bibfnamefont {A.}~\bibnamefont
  {Westphal}},\ }\href {\doibase 10.1103/PhysRevD.82.046003} {\bibfield
  {journal} {\bibinfo  {journal} {Phys.Rev.}\ }\textbf {\bibinfo {volume}
  {D82}},\ \bibinfo {pages} {046003} (\bibinfo {year} {2010})},\ \Eprint
  {http://arxiv.org/abs/0808.0706} {arXiv:0808.0706 [hep-th]} \BibitemShut
  {NoStop}%
\bibitem [{\citenamefont {Kaloper}\ and\ \citenamefont
  {Sorbo}(2009)}]{Kaloper:2008fb}%
  \BibitemOpen
  \bibfield  {author} {\bibinfo {author} {\bibfnamefont {N.}~\bibnamefont
  {Kaloper}}\ and\ \bibinfo {author} {\bibfnamefont {L.}~\bibnamefont
  {Sorbo}},\ }\href {\doibase 10.1103/PhysRevLett.102.121301} {\bibfield
  {journal} {\bibinfo  {journal} {Phys.Rev.Lett.}\ }\textbf {\bibinfo {volume}
  {102}},\ \bibinfo {pages} {121301} (\bibinfo {year} {2009})},\ \Eprint
  {http://arxiv.org/abs/0811.1989} {arXiv:0811.1989 [hep-th]} \BibitemShut
  {NoStop}%
\bibitem [{\citenamefont {Ashoorioon}\ \emph {et~al.}(2009)\citenamefont
  {Ashoorioon}, \citenamefont {Firouzjahi},\ and\ \citenamefont
  {Sheikh-Jabbari}}]{Ashoorioon:2009wa}%
  \BibitemOpen
  \bibfield  {author} {\bibinfo {author} {\bibfnamefont {A.}~\bibnamefont
  {Ashoorioon}}, \bibinfo {author} {\bibfnamefont {H.}~\bibnamefont
  {Firouzjahi}}, \ and\ \bibinfo {author} {\bibfnamefont {M.}~\bibnamefont
  {Sheikh-Jabbari}},\ }\href {\doibase 10.1088/1475-7516/2009/06/018}
  {\bibfield  {journal} {\bibinfo  {journal} {JCAP}\ }\textbf {\bibinfo
  {volume} {0906}},\ \bibinfo {pages} {018} (\bibinfo {year} {2009})},\ \Eprint
  {http://arxiv.org/abs/0903.1481} {arXiv:0903.1481 [hep-th]} \BibitemShut
  {NoStop}%
\bibitem [{\citenamefont {Kaloper}\ \emph {et~al.}(2011)\citenamefont
  {Kaloper}, \citenamefont {Lawrence},\ and\ \citenamefont
  {Sorbo}}]{Kaloper:2011jz}%
  \BibitemOpen
  \bibfield  {author} {\bibinfo {author} {\bibfnamefont {N.}~\bibnamefont
  {Kaloper}}, \bibinfo {author} {\bibfnamefont {A.}~\bibnamefont {Lawrence}}, \
  and\ \bibinfo {author} {\bibfnamefont {L.}~\bibnamefont {Sorbo}},\ }\href
  {\doibase 10.1088/1475-7516/2011/03/023} {\bibfield  {journal} {\bibinfo
  {journal} {JCAP}\ }\textbf {\bibinfo {volume} {1103}},\ \bibinfo {pages}
  {023} (\bibinfo {year} {2011})},\ \Eprint {http://arxiv.org/abs/1101.0026}
  {arXiv:1101.0026 [hep-th]} \BibitemShut {NoStop}%
\bibitem [{\citenamefont {Palti}\ and\ \citenamefont
  {Weigand}(2014)}]{Palti:2014kza}%
  \BibitemOpen
  \bibfield  {author} {\bibinfo {author} {\bibfnamefont {E.}~\bibnamefont
  {Palti}}\ and\ \bibinfo {author} {\bibfnamefont {T.}~\bibnamefont
  {Weigand}},\ }\href {\doibase 10.1007/JHEP04(2014)155} {\bibfield  {journal}
  {\bibinfo  {journal} {JHEP}\ }\textbf {\bibinfo {volume} {1404}},\ \bibinfo
  {pages} {155} (\bibinfo {year} {2014})},\ \Eprint
  {http://arxiv.org/abs/1403.7507} {arXiv:1403.7507 [hep-th]} \BibitemShut
  {NoStop}%
\bibitem [{\citenamefont {Marchesano}\ \emph {et~al.}(2014)\citenamefont
  {Marchesano}, \citenamefont {Shiu},\ and\ \citenamefont
  {Uranga}}]{Marchesano:2014mla}%
  \BibitemOpen
  \bibfield  {author} {\bibinfo {author} {\bibfnamefont {F.}~\bibnamefont
  {Marchesano}}, \bibinfo {author} {\bibfnamefont {G.}~\bibnamefont {Shiu}}, \
  and\ \bibinfo {author} {\bibfnamefont {A.~M.}\ \bibnamefont {Uranga}},\
  }\href@noop {} {\  (\bibinfo {year} {2014})},\ \Eprint
  {http://arxiv.org/abs/1404.3040} {arXiv:1404.3040 [hep-th]} \BibitemShut
  {NoStop}%
\bibitem [{\citenamefont {Hebecker}\ \emph
  {et~al.}(2014{\natexlab{a}})\citenamefont {Hebecker}, \citenamefont {Kraus},\
  and\ \citenamefont {Witkowski}}]{Hebecker:2014eua}%
  \BibitemOpen
  \bibfield  {author} {\bibinfo {author} {\bibfnamefont {A.}~\bibnamefont
  {Hebecker}}, \bibinfo {author} {\bibfnamefont {S.~C.}\ \bibnamefont {Kraus}},
  \ and\ \bibinfo {author} {\bibfnamefont {L.~T.}\ \bibnamefont {Witkowski}},\
  }\href {\doibase 10.1016/j.physletb.2014.08.028} {\bibfield  {journal}
  {\bibinfo  {journal} {Phys.Lett.}\ }\textbf {\bibinfo {volume} {B737}},\
  \bibinfo {pages} {16} (\bibinfo {year} {2014}{\natexlab{a}})},\ \Eprint
  {http://arxiv.org/abs/1404.3711} {arXiv:1404.3711 [hep-th]} \BibitemShut
  {NoStop}%
\bibitem [{\citenamefont {Arends}\ \emph {et~al.}(2014)\citenamefont {Arends},
  \citenamefont {Hebecker}, \citenamefont {Heimpel}, \citenamefont {Kraus},
  \citenamefont {Lust} \emph {et~al.}}]{Arends:2014qca}%
  \BibitemOpen
  \bibfield  {author} {\bibinfo {author} {\bibfnamefont {M.}~\bibnamefont
  {Arends}}, \bibinfo {author} {\bibfnamefont {A.}~\bibnamefont {Hebecker}},
  \bibinfo {author} {\bibfnamefont {K.}~\bibnamefont {Heimpel}}, \bibinfo
  {author} {\bibfnamefont {S.~C.}\ \bibnamefont {Kraus}}, \bibinfo {author}
  {\bibfnamefont {D.}~\bibnamefont {Lust}},  \emph {et~al.},\ }\href {\doibase
  10.1002/prop.201400045} {\bibfield  {journal} {\bibinfo  {journal}
  {Fortsch.Phys.}\ }\textbf {\bibinfo {volume} {62}},\ \bibinfo {pages} {647}
  (\bibinfo {year} {2014})},\ \Eprint {http://arxiv.org/abs/1405.0283}
  {arXiv:1405.0283 [hep-th]} \BibitemShut {NoStop}%
\bibitem [{\citenamefont {McAllister}\ \emph {et~al.}(2014)\citenamefont
  {McAllister}, \citenamefont {Silverstein}, \citenamefont {Westphal},\ and\
  \citenamefont {Wrase}}]{McAllister:2014mpa}%
  \BibitemOpen
  \bibfield  {author} {\bibinfo {author} {\bibfnamefont {L.}~\bibnamefont
  {McAllister}}, \bibinfo {author} {\bibfnamefont {E.}~\bibnamefont
  {Silverstein}}, \bibinfo {author} {\bibfnamefont {A.}~\bibnamefont
  {Westphal}}, \ and\ \bibinfo {author} {\bibfnamefont {T.}~\bibnamefont
  {Wrase}},\ }\href {\doibase 10.1007/JHEP09(2014)123} {\bibfield  {journal}
  {\bibinfo  {journal} {JHEP}\ }\textbf {\bibinfo {volume} {1409}},\ \bibinfo
  {pages} {123} (\bibinfo {year} {2014})},\ \Eprint
  {http://arxiv.org/abs/1405.3652} {arXiv:1405.3652 [hep-th]} \BibitemShut
  {NoStop}%
\bibitem [{\citenamefont {Baumann}\ and\ \citenamefont
  {McAllister}(2014)}]{Baumann:2014nda}%
  \BibitemOpen
  \bibfield  {author} {\bibinfo {author} {\bibfnamefont {D.}~\bibnamefont
  {Baumann}}\ and\ \bibinfo {author} {\bibfnamefont {L.}~\bibnamefont
  {McAllister}},\ }\href@noop {} {\  (\bibinfo {year} {2014})},\ \Eprint
  {http://arxiv.org/abs/1404.2601} {arXiv:1404.2601 [hep-th]} \BibitemShut
  {NoStop}%
\bibitem [{\citenamefont {Banks}\ \emph {et~al.}(2003)\citenamefont {Banks},
  \citenamefont {Dine}, \citenamefont {Fox},\ and\ \citenamefont
  {Gorbatov}}]{Banks:2003sx}%
  \BibitemOpen
  \bibfield  {author} {\bibinfo {author} {\bibfnamefont {T.}~\bibnamefont
  {Banks}}, \bibinfo {author} {\bibfnamefont {M.}~\bibnamefont {Dine}},
  \bibinfo {author} {\bibfnamefont {P.~J.}\ \bibnamefont {Fox}}, \ and\
  \bibinfo {author} {\bibfnamefont {E.}~\bibnamefont {Gorbatov}},\ }\href
  {\doibase 10.1088/1475-7516/2003/06/001} {\bibfield  {journal} {\bibinfo
  {journal} {JCAP}\ }\textbf {\bibinfo {volume} {0306}},\ \bibinfo {pages}
  {001} (\bibinfo {year} {2003})},\ \Eprint
  {http://arxiv.org/abs/hep-th/0303252} {arXiv:hep-th/0303252 [hep-th]}
  \BibitemShut {NoStop}%
\bibitem [{\citenamefont {Svrcek}\ and\ \citenamefont
  {Witten}(2006)}]{Svrcek:2006yi}%
  \BibitemOpen
  \bibfield  {author} {\bibinfo {author} {\bibfnamefont {P.}~\bibnamefont
  {Svrcek}}\ and\ \bibinfo {author} {\bibfnamefont {E.}~\bibnamefont
  {Witten}},\ }\href {\doibase 10.1088/1126-6708/2006/06/051} {\bibfield
  {journal} {\bibinfo  {journal} {JHEP}\ }\textbf {\bibinfo {volume} {0606}},\
  \bibinfo {pages} {051} (\bibinfo {year} {2006})},\ \Eprint
  {http://arxiv.org/abs/hep-th/0605206} {arXiv:hep-th/0605206 [hep-th]}
  \BibitemShut {NoStop}%
\bibitem [{\citenamefont {Rudelius}(2014)}]{Rudelius:2014wla}%
  \BibitemOpen
  \bibfield  {author} {\bibinfo {author} {\bibfnamefont {T.}~\bibnamefont
  {Rudelius}},\ }\href@noop {} {\  (\bibinfo {year} {2014})},\ \Eprint
  {http://arxiv.org/abs/1409.5793} {arXiv:1409.5793 [hep-th]} \BibitemShut
  {NoStop}%
\bibitem [{\citenamefont {Grimm}(2014)}]{Grimm:2014vva}%
  \BibitemOpen
  \bibfield  {author} {\bibinfo {author} {\bibfnamefont {T.~W.}\ \bibnamefont
  {Grimm}},\ }\href@noop {} {\  (\bibinfo {year} {2014})},\ \Eprint
  {http://arxiv.org/abs/1404.4268} {arXiv:1404.4268 [hep-th]} \BibitemShut
  {NoStop}%
\bibitem [{\citenamefont {Kenton}\ and\ \citenamefont
  {Thomas}(2014)}]{Kenton:2014gma}%
  \BibitemOpen
  \bibfield  {author} {\bibinfo {author} {\bibfnamefont {Z.}~\bibnamefont
  {Kenton}}\ and\ \bibinfo {author} {\bibfnamefont {S.}~\bibnamefont
  {Thomas}},\ }\href@noop {} {\  (\bibinfo {year} {2014})},\ \Eprint
  {http://arxiv.org/abs/1409.1221} {arXiv:1409.1221 [hep-th]} \BibitemShut
  {NoStop}%
\bibitem [{\citenamefont {Kim}\ \emph {et~al.}(2005)\citenamefont {Kim},
  \citenamefont {Nilles},\ and\ \citenamefont {Peloso}}]{Kim:2004rp}%
  \BibitemOpen
  \bibfield  {author} {\bibinfo {author} {\bibfnamefont {J.~E.}\ \bibnamefont
  {Kim}}, \bibinfo {author} {\bibfnamefont {H.~P.}\ \bibnamefont {Nilles}}, \
  and\ \bibinfo {author} {\bibfnamefont {M.}~\bibnamefont {Peloso}},\ }\href
  {\doibase 10.1088/1475-7516/2005/01/005} {\bibfield  {journal} {\bibinfo
  {journal} {JCAP}\ }\textbf {\bibinfo {volume} {0501}},\ \bibinfo {pages}
  {005} (\bibinfo {year} {2005})},\ \Eprint
  {http://arxiv.org/abs/hep-ph/0409138} {arXiv:hep-ph/0409138 [hep-ph]}
  \BibitemShut {NoStop}%
\bibitem [{\citenamefont {Czerny}\ \emph
  {et~al.}(2014{\natexlab{b}})\citenamefont {Czerny}, \citenamefont {Higaki},\
  and\ \citenamefont {Takahashi}}]{Czerny:2014qqa}%
  \BibitemOpen
  \bibfield  {author} {\bibinfo {author} {\bibfnamefont {M.}~\bibnamefont
  {Czerny}}, \bibinfo {author} {\bibfnamefont {T.}~\bibnamefont {Higaki}}, \
  and\ \bibinfo {author} {\bibfnamefont {F.}~\bibnamefont {Takahashi}},\ }\href
  {\doibase 10.1016/j.physletb.2014.05.041} {\bibfield  {journal} {\bibinfo
  {journal} {Phys.Lett.}\ }\textbf {\bibinfo {volume} {B734}},\ \bibinfo
  {pages} {167} (\bibinfo {year} {2014}{\natexlab{b}})},\ \Eprint
  {http://arxiv.org/abs/1403.5883} {arXiv:1403.5883 [hep-ph]} \BibitemShut
  {NoStop}%
\bibitem [{\citenamefont {Harigaya}\ and\ \citenamefont
  {Ibe}(2014)}]{Harigaya:2014eta}%
  \BibitemOpen
  \bibfield  {author} {\bibinfo {author} {\bibfnamefont {K.}~\bibnamefont
  {Harigaya}}\ and\ \bibinfo {author} {\bibfnamefont {M.}~\bibnamefont {Ibe}},\
  }\href@noop {} {\  (\bibinfo {year} {2014})},\ \Eprint
  {http://arxiv.org/abs/1404.3511} {arXiv:1404.3511 [hep-ph]} \BibitemShut
  {NoStop}%
\bibitem [{\citenamefont {Choi}\ \emph {et~al.}(2014)\citenamefont {Choi},
  \citenamefont {Kim},\ and\ \citenamefont {Yun}}]{Choi:2014rja}%
  \BibitemOpen
  \bibfield  {author} {\bibinfo {author} {\bibfnamefont {K.}~\bibnamefont
  {Choi}}, \bibinfo {author} {\bibfnamefont {H.}~\bibnamefont {Kim}}, \ and\
  \bibinfo {author} {\bibfnamefont {S.}~\bibnamefont {Yun}},\ }\href {\doibase
  10.1103/PhysRevD.90.023545} {\bibfield  {journal} {\bibinfo  {journal}
  {Phys.Rev.}\ }\textbf {\bibinfo {volume} {D90}},\ \bibinfo {pages} {023545}
  (\bibinfo {year} {2014})},\ \Eprint {http://arxiv.org/abs/1404.6209}
  {arXiv:1404.6209 [hep-th]} \BibitemShut {NoStop}%
\bibitem [{\citenamefont {Higaki}\ and\ \citenamefont
  {Takahashi}(2014)}]{Higaki:2014pja}%
  \BibitemOpen
  \bibfield  {author} {\bibinfo {author} {\bibfnamefont {T.}~\bibnamefont
  {Higaki}}\ and\ \bibinfo {author} {\bibfnamefont {F.}~\bibnamefont
  {Takahashi}},\ }\href {\doibase 10.1007/JHEP07(2014)074} {\bibfield
  {journal} {\bibinfo  {journal} {JHEP}\ }\textbf {\bibinfo {volume} {1407}},\
  \bibinfo {pages} {074} (\bibinfo {year} {2014})},\ \Eprint
  {http://arxiv.org/abs/1404.6923} {arXiv:1404.6923 [hep-th]} \BibitemShut
  {NoStop}%
\bibitem [{\citenamefont {Tye}\ and\ \citenamefont {Wong}(2014)}]{Tye:2014tja}%
  \BibitemOpen
  \bibfield  {author} {\bibinfo {author} {\bibfnamefont {S.~H.~H.}\
  \bibnamefont {Tye}}\ and\ \bibinfo {author} {\bibfnamefont {S.~S.~C.}\
  \bibnamefont {Wong}},\ }\href@noop {} {\  (\bibinfo {year} {2014})},\ \Eprint
  {http://arxiv.org/abs/1404.6988} {arXiv:1404.6988 [astro-ph.CO]} \BibitemShut
  {NoStop}%
\bibitem [{\citenamefont {Ben-Dayan}\ \emph
  {et~al.}(2014{\natexlab{a}})\citenamefont {Ben-Dayan}, \citenamefont
  {Pedro},\ and\ \citenamefont {Westphal}}]{Ben-Dayan:2014zsa}%
  \BibitemOpen
  \bibfield  {author} {\bibinfo {author} {\bibfnamefont {I.}~\bibnamefont
  {Ben-Dayan}}, \bibinfo {author} {\bibfnamefont {F.~G.}\ \bibnamefont
  {Pedro}}, \ and\ \bibinfo {author} {\bibfnamefont {A.}~\bibnamefont
  {Westphal}},\ }\href@noop {} {\  (\bibinfo {year} {2014}{\natexlab{a}})},\
  \Eprint {http://arxiv.org/abs/1404.7773} {arXiv:1404.7773 [hep-th]}
  \BibitemShut {NoStop}%
\bibitem [{\citenamefont {Kappl}\ \emph {et~al.}(2014)\citenamefont {Kappl},
  \citenamefont {Krippendorf},\ and\ \citenamefont {Nilles}}]{Kappl:2014lra}%
  \BibitemOpen
  \bibfield  {author} {\bibinfo {author} {\bibfnamefont {R.}~\bibnamefont
  {Kappl}}, \bibinfo {author} {\bibfnamefont {S.}~\bibnamefont {Krippendorf}},
  \ and\ \bibinfo {author} {\bibfnamefont {H.~P.}\ \bibnamefont {Nilles}},\
  }\href {\doibase 10.1016/j.physletb.2014.08.045} {\bibfield  {journal}
  {\bibinfo  {journal} {Phys.Lett.}\ }\textbf {\bibinfo {volume} {B737}},\
  \bibinfo {pages} {124} (\bibinfo {year} {2014})},\ \Eprint
  {http://arxiv.org/abs/1404.7127} {arXiv:1404.7127 [hep-th]} \BibitemShut
  {NoStop}%
\bibitem [{\citenamefont {Long}\ \emph {et~al.}(2014)\citenamefont {Long},
  \citenamefont {McAllister},\ and\ \citenamefont {McGuirk}}]{Long:2014dta}%
  \BibitemOpen
  \bibfield  {author} {\bibinfo {author} {\bibfnamefont {C.}~\bibnamefont
  {Long}}, \bibinfo {author} {\bibfnamefont {L.}~\bibnamefont {McAllister}}, \
  and\ \bibinfo {author} {\bibfnamefont {P.}~\bibnamefont {McGuirk}},\ }\href
  {\doibase 10.1103/PhysRevD.90.023501} {\bibfield  {journal} {\bibinfo
  {journal} {Phys.Rev.}\ }\textbf {\bibinfo {volume} {D90}},\ \bibinfo {pages}
  {023501} (\bibinfo {year} {2014})},\ \Eprint {http://arxiv.org/abs/1404.7852}
  {arXiv:1404.7852 [hep-th]} \BibitemShut {NoStop}%
\bibitem [{\citenamefont {Ben-Dayan}\ \emph
  {et~al.}(2014{\natexlab{b}})\citenamefont {Ben-Dayan}, \citenamefont
  {Pedro},\ and\ \citenamefont {Westphal}}]{Ben-Dayan:2014lca}%
  \BibitemOpen
  \bibfield  {author} {\bibinfo {author} {\bibfnamefont {I.}~\bibnamefont
  {Ben-Dayan}}, \bibinfo {author} {\bibfnamefont {F.~G.}\ \bibnamefont
  {Pedro}}, \ and\ \bibinfo {author} {\bibfnamefont {A.}~\bibnamefont
  {Westphal}},\ }\href@noop {} {\  (\bibinfo {year} {2014}{\natexlab{b}})},\
  \Eprint {http://arxiv.org/abs/1407.2562} {arXiv:1407.2562 [hep-th]}
  \BibitemShut {NoStop}%
\bibitem [{\citenamefont {Westphal}(2014)}]{Westphal:2014ana}%
  \BibitemOpen
  \bibfield  {author} {\bibinfo {author} {\bibfnamefont {A.}~\bibnamefont
  {Westphal}},\ }\href@noop {} {\  (\bibinfo {year} {2014})},\ \Eprint
  {http://arxiv.org/abs/1409.5350} {arXiv:1409.5350 [hep-th]} \BibitemShut
  {NoStop}%
\bibitem [{\citenamefont {Gao}\ \emph {et~al.}(2014)\citenamefont {Gao},
  \citenamefont {Li},\ and\ \citenamefont {Shukla}}]{Gao:2014uha}%
  \BibitemOpen
  \bibfield  {author} {\bibinfo {author} {\bibfnamefont {X.}~\bibnamefont
  {Gao}}, \bibinfo {author} {\bibfnamefont {T.}~\bibnamefont {Li}}, \ and\
  \bibinfo {author} {\bibfnamefont {P.}~\bibnamefont {Shukla}},\ }\href@noop {}
  {\  (\bibinfo {year} {2014})},\ \Eprint {http://arxiv.org/abs/1406.0341}
  {arXiv:1406.0341 [hep-th]} \BibitemShut {NoStop}%
\bibitem [{\citenamefont {Coleman}\ and\ \citenamefont
  {De~Luccia}(1980)}]{Coleman:1980aw}%
  \BibitemOpen
  \bibfield  {author} {\bibinfo {author} {\bibfnamefont {S.~R.}\ \bibnamefont
  {Coleman}}\ and\ \bibinfo {author} {\bibfnamefont {F.}~\bibnamefont
  {De~Luccia}},\ }\href {\doibase 10.1103/PhysRevD.21.3305} {\bibfield
  {journal} {\bibinfo  {journal} {Phys.Rev.}\ }\textbf {\bibinfo {volume}
  {D21}},\ \bibinfo {pages} {3305} (\bibinfo {year} {1980})}\BibitemShut
  {NoStop}%
\bibitem [{\citenamefont {Bousso}\ and\ \citenamefont
  {Polchinski}(2000)}]{Bousso:2000xa}%
  \BibitemOpen
  \bibfield  {author} {\bibinfo {author} {\bibfnamefont {R.}~\bibnamefont
  {Bousso}}\ and\ \bibinfo {author} {\bibfnamefont {J.}~\bibnamefont
  {Polchinski}},\ }\href {\doibase 10.1088/1126-6708/2000/06/006} {\bibfield
  {journal} {\bibinfo  {journal} {JHEP}\ }\textbf {\bibinfo {volume} {0006}},\
  \bibinfo {pages} {006} (\bibinfo {year} {2000})},\ \Eprint
  {http://arxiv.org/abs/hep-th/0004134} {arXiv:hep-th/0004134 [hep-th]}
  \BibitemShut {NoStop}%
\bibitem [{\citenamefont {Susskind}(2003)}]{Susskind:2003kw}%
  \BibitemOpen
  \bibfield  {author} {\bibinfo {author} {\bibfnamefont {L.}~\bibnamefont
  {Susskind}},\ }\href@noop {} {\  (\bibinfo {year} {2003})},\ \Eprint
  {http://arxiv.org/abs/hep-th/0302219} {arXiv:hep-th/0302219 [hep-th]}
  \BibitemShut {NoStop}%
\bibitem [{\citenamefont {Linde}(1995)}]{Linde:1995xm}%
  \BibitemOpen
  \bibfield  {author} {\bibinfo {author} {\bibfnamefont {A.~D.}\ \bibnamefont
  {Linde}},\ }\href {\doibase 10.1016/0370-2693(95)00370-Z} {\bibfield
  {journal} {\bibinfo  {journal} {Phys.Lett.}\ }\textbf {\bibinfo {volume}
  {B351}},\ \bibinfo {pages} {99} (\bibinfo {year} {1995})},\ \Eprint
  {http://arxiv.org/abs/hep-th/9503097} {arXiv:hep-th/9503097 [hep-th]}
  \BibitemShut {NoStop}%
\bibitem [{\citenamefont {Linde}\ and\ \citenamefont
  {Mezhlumian}(1995)}]{Linde:1995rv}%
  \BibitemOpen
  \bibfield  {author} {\bibinfo {author} {\bibfnamefont {A.~D.}\ \bibnamefont
  {Linde}}\ and\ \bibinfo {author} {\bibfnamefont {A.}~\bibnamefont
  {Mezhlumian}},\ }\href {\doibase 10.1103/PhysRevD.52.6789} {\bibfield
  {journal} {\bibinfo  {journal} {Phys.Rev.}\ }\textbf {\bibinfo {volume}
  {D52}},\ \bibinfo {pages} {6789} (\bibinfo {year} {1995})},\ \Eprint
  {http://arxiv.org/abs/astro-ph/9506017} {arXiv:astro-ph/9506017 [astro-ph]}
  \BibitemShut {NoStop}%
\bibitem [{\citenamefont {Freivogel}\ \emph {et~al.}(2006)\citenamefont
  {Freivogel}, \citenamefont {Kleban}, \citenamefont {Rodriguez~Martinez},\
  and\ \citenamefont {Susskind}}]{Freivogel:2005vv}%
  \BibitemOpen
  \bibfield  {author} {\bibinfo {author} {\bibfnamefont {B.}~\bibnamefont
  {Freivogel}}, \bibinfo {author} {\bibfnamefont {M.}~\bibnamefont {Kleban}},
  \bibinfo {author} {\bibfnamefont {M.}~\bibnamefont {Rodriguez~Martinez}}, \
  and\ \bibinfo {author} {\bibfnamefont {L.}~\bibnamefont {Susskind}},\ }\href
  {\doibase 10.1088/1126-6708/2006/03/039} {\bibfield  {journal} {\bibinfo
  {journal} {JHEP}\ }\textbf {\bibinfo {volume} {0603}},\ \bibinfo {pages}
  {039} (\bibinfo {year} {2006})},\ \Eprint
  {http://arxiv.org/abs/hep-th/0505232} {arXiv:hep-th/0505232 [hep-th]}
  \BibitemShut {NoStop}%
\bibitem [{\citenamefont {Sugimura}\ \emph {et~al.}(2012)\citenamefont
  {Sugimura}, \citenamefont {Yamauchi},\ and\ \citenamefont
  {Sasaki}}]{Sugimura:2011tk}%
  \BibitemOpen
  \bibfield  {author} {\bibinfo {author} {\bibfnamefont {K.}~\bibnamefont
  {Sugimura}}, \bibinfo {author} {\bibfnamefont {D.}~\bibnamefont {Yamauchi}},
  \ and\ \bibinfo {author} {\bibfnamefont {M.}~\bibnamefont {Sasaki}},\ }\href
  {\doibase 10.1088/1475-7516/2012/01/027} {\bibfield  {journal} {\bibinfo
  {journal} {JCAP}\ }\textbf {\bibinfo {volume} {1201}},\ \bibinfo {pages}
  {027} (\bibinfo {year} {2012})},\ \Eprint {http://arxiv.org/abs/1110.4773}
  {arXiv:1110.4773 [gr-qc]} \BibitemShut {NoStop}%
\bibitem [{\citenamefont {Freivogel}\ \emph {et~al.}(2014)\citenamefont
  {Freivogel}, \citenamefont {Kleban}, \citenamefont {Martinez},\ and\
  \citenamefont {Susskind}}]{Freivogel:2014hca}%
  \BibitemOpen
  \bibfield  {author} {\bibinfo {author} {\bibfnamefont {B.}~\bibnamefont
  {Freivogel}}, \bibinfo {author} {\bibfnamefont {M.}~\bibnamefont {Kleban}},
  \bibinfo {author} {\bibfnamefont {M.~R.}\ \bibnamefont {Martinez}}, \ and\
  \bibinfo {author} {\bibfnamefont {L.}~\bibnamefont {Susskind}},\ }\href@noop
  {} {\  (\bibinfo {year} {2014})},\ \Eprint {http://arxiv.org/abs/1404.2274}
  {arXiv:1404.2274 [astro-ph.CO]} \BibitemShut {NoStop}%
\bibitem [{\citenamefont {Yamauchi}\ \emph {et~al.}(2011)\citenamefont
  {Yamauchi}, \citenamefont {Linde}, \citenamefont {Naruko}, \citenamefont
  {Sasaki},\ and\ \citenamefont {Tanaka}}]{Yamauchi:2011qq}%
  \BibitemOpen
  \bibfield  {author} {\bibinfo {author} {\bibfnamefont {D.}~\bibnamefont
  {Yamauchi}}, \bibinfo {author} {\bibfnamefont {A.}~\bibnamefont {Linde}},
  \bibinfo {author} {\bibfnamefont {A.}~\bibnamefont {Naruko}}, \bibinfo
  {author} {\bibfnamefont {M.}~\bibnamefont {Sasaki}}, \ and\ \bibinfo {author}
  {\bibfnamefont {T.}~\bibnamefont {Tanaka}},\ }\href {\doibase
  10.1103/PhysRevD.84.043513} {\bibfield  {journal} {\bibinfo  {journal}
  {Phys.Rev.}\ }\textbf {\bibinfo {volume} {D84}},\ \bibinfo {pages} {043513}
  (\bibinfo {year} {2011})},\ \Eprint {http://arxiv.org/abs/1105.2674}
  {arXiv:1105.2674 [hep-th]} \BibitemShut {NoStop}%
\bibitem [{\citenamefont {Bousso}\ \emph {et~al.}(2013)\citenamefont {Bousso},
  \citenamefont {Harlow},\ and\ \citenamefont {Senatore}}]{Bousso:2013uia}%
  \BibitemOpen
  \bibfield  {author} {\bibinfo {author} {\bibfnamefont {R.}~\bibnamefont
  {Bousso}}, \bibinfo {author} {\bibfnamefont {D.}~\bibnamefont {Harlow}}, \
  and\ \bibinfo {author} {\bibfnamefont {L.}~\bibnamefont {Senatore}},\
  }\href@noop {} {\  (\bibinfo {year} {2013})},\ \Eprint
  {http://arxiv.org/abs/1309.4060} {arXiv:1309.4060 [hep-th]} \BibitemShut
  {NoStop}%
\bibitem [{\citenamefont {Bousso}\ \emph {et~al.}(2014)\citenamefont {Bousso},
  \citenamefont {Harlow},\ and\ \citenamefont {Senatore}}]{Bousso:2014jca}%
  \BibitemOpen
  \bibfield  {author} {\bibinfo {author} {\bibfnamefont {R.}~\bibnamefont
  {Bousso}}, \bibinfo {author} {\bibfnamefont {D.}~\bibnamefont {Harlow}}, \
  and\ \bibinfo {author} {\bibfnamefont {L.}~\bibnamefont {Senatore}},\
  }\href@noop {} {\  (\bibinfo {year} {2014})},\ \Eprint
  {http://arxiv.org/abs/1404.2278} {arXiv:1404.2278 [astro-ph.CO]} \BibitemShut
  {NoStop}%
\bibitem [{\citenamefont {Murayama}\ \emph {et~al.}(2014)\citenamefont
  {Murayama}, \citenamefont {Nakayama}, \citenamefont {Takahashi},\ and\
  \citenamefont {Yanagida}}]{Murayama:2014saa}%
  \BibitemOpen
  \bibfield  {author} {\bibinfo {author} {\bibfnamefont {H.}~\bibnamefont
  {Murayama}}, \bibinfo {author} {\bibfnamefont {K.}~\bibnamefont {Nakayama}},
  \bibinfo {author} {\bibfnamefont {F.}~\bibnamefont {Takahashi}}, \ and\
  \bibinfo {author} {\bibfnamefont {T.~T.}\ \bibnamefont {Yanagida}},\
  }\href@noop {} {\  (\bibinfo {year} {2014})},\ \Eprint
  {http://arxiv.org/abs/1404.3857} {arXiv:1404.3857 [hep-ph]} \BibitemShut
  {NoStop}%
\bibitem [{\citenamefont {Kobayashi}\ and\ \citenamefont
  {Takahashi}(2011)}]{Kobayashi:2010pz}%
  \BibitemOpen
  \bibfield  {author} {\bibinfo {author} {\bibfnamefont {T.}~\bibnamefont
  {Kobayashi}}\ and\ \bibinfo {author} {\bibfnamefont {F.}~\bibnamefont
  {Takahashi}},\ }\href {\doibase 10.1088/1475-7516/2011/01/026} {\bibfield
  {journal} {\bibinfo  {journal} {JCAP}\ }\textbf {\bibinfo {volume} {1101}},\
  \bibinfo {pages} {026} (\bibinfo {year} {2011})},\ \Eprint
  {http://arxiv.org/abs/1011.3988} {arXiv:1011.3988 [astro-ph.CO]} \BibitemShut
  {NoStop}%
\bibitem [{\citenamefont {Takahashi}(2013)}]{Takahashi:2013tj}%
  \BibitemOpen
  \bibfield  {author} {\bibinfo {author} {\bibfnamefont {F.}~\bibnamefont
  {Takahashi}},\ }\href {\doibase 10.1088/1475-7516/2013/06/013} {\bibfield
  {journal} {\bibinfo  {journal} {JCAP}\ }\textbf {\bibinfo {volume} {1306}},\
  \bibinfo {pages} {013} (\bibinfo {year} {2013})},\ \Eprint
  {http://arxiv.org/abs/1301.2834} {arXiv:1301.2834} \BibitemShut {NoStop}%
\bibitem [{\citenamefont {Czerny}\ \emph
  {et~al.}(2014{\natexlab{c}})\citenamefont {Czerny}, \citenamefont
  {Kobayashi},\ and\ \citenamefont {Takahashi}}]{Czerny:2014wua}%
  \BibitemOpen
  \bibfield  {author} {\bibinfo {author} {\bibfnamefont {M.}~\bibnamefont
  {Czerny}}, \bibinfo {author} {\bibfnamefont {T.}~\bibnamefont {Kobayashi}}, \
  and\ \bibinfo {author} {\bibfnamefont {F.}~\bibnamefont {Takahashi}},\
  }\href@noop {} {\  (\bibinfo {year} {2014}{\natexlab{c}})},\ \Eprint
  {http://arxiv.org/abs/1403.4589} {arXiv:1403.4589 [astro-ph.CO]} \BibitemShut
  {NoStop}%
\bibitem [{\citenamefont {Abazajian}\ \emph {et~al.}(2014)\citenamefont
  {Abazajian}, \citenamefont {Aslanyan}, \citenamefont {Easther},\ and\
  \citenamefont {Price}}]{Abazajian:2014tqa}%
  \BibitemOpen
  \bibfield  {author} {\bibinfo {author} {\bibfnamefont {K.~N.}\ \bibnamefont
  {Abazajian}}, \bibinfo {author} {\bibfnamefont {G.}~\bibnamefont {Aslanyan}},
  \bibinfo {author} {\bibfnamefont {R.}~\bibnamefont {Easther}}, \ and\
  \bibinfo {author} {\bibfnamefont {L.~C.}\ \bibnamefont {Price}},\ }\href
  {\doibase 10.1088/1475-7516/2014/08/053} {\bibfield  {journal} {\bibinfo
  {journal} {JCAP}\ }\textbf {\bibinfo {volume} {1408}},\ \bibinfo {pages}
  {053} (\bibinfo {year} {2014})},\ \Eprint {http://arxiv.org/abs/1403.5922}
  {arXiv:1403.5922 [astro-ph.CO]} \BibitemShut {NoStop}%
\bibitem [{\citenamefont {Garrison-Kimmel}\ \emph {et~al.}(2014)\citenamefont
  {Garrison-Kimmel}, \citenamefont {Horiuchi}, \citenamefont {Abazajian},
  \citenamefont {Bullock},\ and\ \citenamefont
  {Kaplinghat}}]{Garrison-Kimmel:2014kia}%
  \BibitemOpen
  \bibfield  {author} {\bibinfo {author} {\bibfnamefont {S.}~\bibnamefont
  {Garrison-Kimmel}}, \bibinfo {author} {\bibfnamefont {S.}~\bibnamefont
  {Horiuchi}}, \bibinfo {author} {\bibfnamefont {K.~N.}\ \bibnamefont
  {Abazajian}}, \bibinfo {author} {\bibfnamefont {J.~S.}\ \bibnamefont
  {Bullock}}, \ and\ \bibinfo {author} {\bibfnamefont {M.}~\bibnamefont
  {Kaplinghat}},\ }\href {\doibase 10.1093/mnras/stu1479} {\bibfield  {journal}
  {\bibinfo  {journal} {Mon.Not.Roy.Astron.Soc.}\ }\textbf {\bibinfo {volume}
  {444}},\ \bibinfo {pages} {961} (\bibinfo {year} {2014})},\ \Eprint
  {http://arxiv.org/abs/1405.3985} {arXiv:1405.3985 [astro-ph.CO]} \BibitemShut
  {NoStop}%
\bibitem [{\citenamefont {Fukugita}\ and\ \citenamefont
  {Yanagida}(1986)}]{Fukugita:1986hr}%
  \BibitemOpen
  \bibfield  {author} {\bibinfo {author} {\bibfnamefont {M.}~\bibnamefont
  {Fukugita}}\ and\ \bibinfo {author} {\bibfnamefont {T.}~\bibnamefont
  {Yanagida}},\ }\href {\doibase 10.1016/0370-2693(86)91126-3} {\bibfield
  {journal} {\bibinfo  {journal} {Phys.Lett.}\ }\textbf {\bibinfo {volume}
  {B174}},\ \bibinfo {pages} {45} (\bibinfo {year} {1986})}\BibitemShut
  {NoStop}%
\bibitem [{\citenamefont {Higaki}\ \emph
  {et~al.}(2014{\natexlab{a}})\citenamefont {Higaki}, \citenamefont
  {Kitajima},\ and\ \citenamefont {Takahashi}}]{Higaki:2014qua}%
  \BibitemOpen
  \bibfield  {author} {\bibinfo {author} {\bibfnamefont {T.}~\bibnamefont
  {Higaki}}, \bibinfo {author} {\bibfnamefont {N.}~\bibnamefont {Kitajima}}, \
  and\ \bibinfo {author} {\bibfnamefont {F.}~\bibnamefont {Takahashi}},\
  }\href@noop {} {\  (\bibinfo {year} {2014}{\natexlab{a}})},\ \Eprint
  {http://arxiv.org/abs/1408.3936} {arXiv:1408.3936 [hep-ph]} \BibitemShut
  {NoStop}%
\bibitem [{\citenamefont {Barnaby}\ and\ \citenamefont
  {Peloso}(2011)}]{Barnaby:2010vf}%
  \BibitemOpen
  \bibfield  {author} {\bibinfo {author} {\bibfnamefont {N.}~\bibnamefont
  {Barnaby}}\ and\ \bibinfo {author} {\bibfnamefont {M.}~\bibnamefont
  {Peloso}},\ }\href {\doibase 10.1103/PhysRevLett.106.181301} {\bibfield
  {journal} {\bibinfo  {journal} {Phys.Rev.Lett.}\ }\textbf {\bibinfo {volume}
  {106}},\ \bibinfo {pages} {181301} (\bibinfo {year} {2011})},\ \Eprint
  {http://arxiv.org/abs/1011.1500} {arXiv:1011.1500 [hep-ph]} \BibitemShut
  {NoStop}%
\bibitem [{\citenamefont {Barnaby}\ \emph {et~al.}(2011)\citenamefont
  {Barnaby}, \citenamefont {Namba},\ and\ \citenamefont
  {Peloso}}]{Barnaby:2011vw}%
  \BibitemOpen
  \bibfield  {author} {\bibinfo {author} {\bibfnamefont {N.}~\bibnamefont
  {Barnaby}}, \bibinfo {author} {\bibfnamefont {R.}~\bibnamefont {Namba}}, \
  and\ \bibinfo {author} {\bibfnamefont {M.}~\bibnamefont {Peloso}},\ }\href
  {\doibase 10.1088/1475-7516/2011/04/009} {\bibfield  {journal} {\bibinfo
  {journal} {JCAP}\ }\textbf {\bibinfo {volume} {1104}},\ \bibinfo {pages}
  {009} (\bibinfo {year} {2011})},\ \Eprint {http://arxiv.org/abs/1102.4333}
  {arXiv:1102.4333 [astro-ph.CO]} \BibitemShut {NoStop}%
\bibitem [{\citenamefont {Kallosh}\ and\ \citenamefont
  {Linde}(2004)}]{Kallosh:2004yh}%
  \BibitemOpen
  \bibfield  {author} {\bibinfo {author} {\bibfnamefont {R.}~\bibnamefont
  {Kallosh}}\ and\ \bibinfo {author} {\bibfnamefont {A.~D.}\ \bibnamefont
  {Linde}},\ }\href {\doibase 10.1088/1126-6708/2004/12/004} {\bibfield
  {journal} {\bibinfo  {journal} {JHEP}\ }\textbf {\bibinfo {volume} {0412}},\
  \bibinfo {pages} {004} (\bibinfo {year} {2004})},\ \Eprint
  {http://arxiv.org/abs/hep-th/0411011} {arXiv:hep-th/0411011 [hep-th]}
  \BibitemShut {NoStop}%
\bibitem [{\citenamefont {Abe}\ \emph {et~al.}(2014)\citenamefont {Abe},
  \citenamefont {Kobayashi},\ and\ \citenamefont {Otsuka}}]{Abe:2014pwa}%
  \BibitemOpen
  \bibfield  {author} {\bibinfo {author} {\bibfnamefont {H.}~\bibnamefont
  {Abe}}, \bibinfo {author} {\bibfnamefont {T.}~\bibnamefont {Kobayashi}}, \
  and\ \bibinfo {author} {\bibfnamefont {H.}~\bibnamefont {Otsuka}},\
  }\href@noop {} {\  (\bibinfo {year} {2014})},\ \Eprint
  {http://arxiv.org/abs/1409.8436} {arXiv:1409.8436 [hep-th]} \BibitemShut
  {NoStop}%
\bibitem [{\citenamefont {Hayashi}\ \emph {et~al.}(2014)\citenamefont
  {Hayashi}, \citenamefont {Matsuda},\ and\ \citenamefont
  {Watari}}]{Hayashi:2014aua}%
  \BibitemOpen
  \bibfield  {author} {\bibinfo {author} {\bibfnamefont {H.}~\bibnamefont
  {Hayashi}}, \bibinfo {author} {\bibfnamefont {R.}~\bibnamefont {Matsuda}}, \
  and\ \bibinfo {author} {\bibfnamefont {T.}~\bibnamefont {Watari}},\
  }\href@noop {} {\  (\bibinfo {year} {2014})},\ \Eprint
  {http://arxiv.org/abs/1410.7522} {arXiv:1410.7522 [hep-th]} \BibitemShut
  {NoStop}%
\bibitem [{\citenamefont {Hebecker}\ \emph
  {et~al.}(2014{\natexlab{b}})\citenamefont {Hebecker}, \citenamefont {Mangat},
  \citenamefont {Rompineve},\ and\ \citenamefont
  {Witkowski}}]{Hebecker:2014kva}%
  \BibitemOpen
  \bibfield  {author} {\bibinfo {author} {\bibfnamefont {A.}~\bibnamefont
  {Hebecker}}, \bibinfo {author} {\bibfnamefont {P.}~\bibnamefont {Mangat}},
  \bibinfo {author} {\bibfnamefont {F.}~\bibnamefont {Rompineve}}, \ and\
  \bibinfo {author} {\bibfnamefont {L.~T.}\ \bibnamefont {Witkowski}},\
  }\href@noop {} {\  (\bibinfo {year} {2014}{\natexlab{b}})},\ \Eprint
  {http://arxiv.org/abs/1411.2032} {arXiv:1411.2032 [hep-th]} \BibitemShut
  {NoStop}%
\bibitem [{\citenamefont {Dine}\ \emph {et~al.}(2014)\citenamefont {Dine},
  \citenamefont {Draper},\ and\ \citenamefont {Monteux}}]{Dine:2014hwa}%
  \BibitemOpen
  \bibfield  {author} {\bibinfo {author} {\bibfnamefont {M.}~\bibnamefont
  {Dine}}, \bibinfo {author} {\bibfnamefont {P.}~\bibnamefont {Draper}}, \ and\
  \bibinfo {author} {\bibfnamefont {A.}~\bibnamefont {Monteux}},\ }\href
  {\doibase 10.1007/JHEP07(2014)146} {\bibfield  {journal} {\bibinfo  {journal}
  {JHEP}\ }\textbf {\bibinfo {volume} {1407}},\ \bibinfo {pages} {146}
  (\bibinfo {year} {2014})},\ \Eprint {http://arxiv.org/abs/1405.0068}
  {arXiv:1405.0068 [hep-th]} \BibitemShut {NoStop}%
\bibitem [{\citenamefont {Yonekura}(2014)}]{Yonekura:2014oja}%
  \BibitemOpen
  \bibfield  {author} {\bibinfo {author} {\bibfnamefont {K.}~\bibnamefont
  {Yonekura}},\ }\href {\doibase 10.1088/1475-7516/2014/10/054} {\bibfield
  {journal} {\bibinfo  {journal} {JCAP}\ }\textbf {\bibinfo {volume} {1410}},\
  \bibinfo {pages} {054} (\bibinfo {year} {2014})},\ \Eprint
  {http://arxiv.org/abs/1405.0734} {arXiv:1405.0734 [hep-th]} \BibitemShut
  {NoStop}%
\bibitem [{\citenamefont {Witten}(1980)}]{Witten:1980sp}%
  \BibitemOpen
  \bibfield  {author} {\bibinfo {author} {\bibfnamefont {E.}~\bibnamefont
  {Witten}},\ }\href {\doibase 10.1016/0003-4916(80)90325-5} {\bibfield
  {journal} {\bibinfo  {journal} {Annals Phys.}\ }\textbf {\bibinfo {volume}
  {128}},\ \bibinfo {pages} {363} (\bibinfo {year} {1980})}\BibitemShut
  {NoStop}%
\bibitem [{\citenamefont {Witten}(1998)}]{Witten:1998uka}%
  \BibitemOpen
  \bibfield  {author} {\bibinfo {author} {\bibfnamefont {E.}~\bibnamefont
  {Witten}},\ }\href {\doibase 10.1103/PhysRevLett.81.2862} {\bibfield
  {journal} {\bibinfo  {journal} {Phys.Rev.Lett.}\ }\textbf {\bibinfo {volume}
  {81}},\ \bibinfo {pages} {2862} (\bibinfo {year} {1998})},\ \Eprint
  {http://arxiv.org/abs/hep-th/9807109} {arXiv:hep-th/9807109 [hep-th]}
  \BibitemShut {NoStop}%
\bibitem [{\citenamefont {Arkani-Hamed}\ \emph {et~al.}(2007)\citenamefont
  {Arkani-Hamed}, \citenamefont {Motl}, \citenamefont {Nicolis},\ and\
  \citenamefont {Vafa}}]{ArkaniHamed:2006dz}%
  \BibitemOpen
  \bibfield  {author} {\bibinfo {author} {\bibfnamefont {N.}~\bibnamefont
  {Arkani-Hamed}}, \bibinfo {author} {\bibfnamefont {L.}~\bibnamefont {Motl}},
  \bibinfo {author} {\bibfnamefont {A.}~\bibnamefont {Nicolis}}, \ and\
  \bibinfo {author} {\bibfnamefont {C.}~\bibnamefont {Vafa}},\ }\href {\doibase
  10.1088/1126-6708/2007/06/060} {\bibfield  {journal} {\bibinfo  {journal}
  {JHEP}\ }\textbf {\bibinfo {volume} {0706}},\ \bibinfo {pages} {060}
  (\bibinfo {year} {2007})},\ \Eprint {http://arxiv.org/abs/hep-th/0601001}
  {arXiv:hep-th/0601001 [hep-th]} \BibitemShut {NoStop}%
\bibitem [{\citenamefont {Pajer}(2008)}]{Pajer:2008uy}%
  \BibitemOpen
  \bibfield  {author} {\bibinfo {author} {\bibfnamefont {E.}~\bibnamefont
  {Pajer}},\ }\href {\doibase 10.1088/1475-7516/2008/04/031} {\bibfield
  {journal} {\bibinfo  {journal} {JCAP}\ }\textbf {\bibinfo {volume} {0804}},\
  \bibinfo {pages} {031} (\bibinfo {year} {2008})},\ \Eprint
  {http://arxiv.org/abs/0802.2916} {arXiv:0802.2916 [hep-th]} \BibitemShut
  {NoStop}%
\bibitem [{\citenamefont {Denef}(2008)}]{Denef:2008wq}%
  \BibitemOpen
  \bibfield  {author} {\bibinfo {author} {\bibfnamefont {F.}~\bibnamefont
  {Denef}},\ }\href@noop {} {\ ,\ \bibinfo {pages} {483} (\bibinfo {year}
  {2008})},\ \Eprint {http://arxiv.org/abs/0803.1194} {arXiv:0803.1194
  [hep-th]} \BibitemShut {NoStop}%
\bibitem [{\citenamefont {Collinucci}\ \emph {et~al.}(2009)\citenamefont
  {Collinucci}, \citenamefont {Denef},\ and\ \citenamefont
  {Esole}}]{Collinucci:2008pf}%
  \BibitemOpen
  \bibfield  {author} {\bibinfo {author} {\bibfnamefont {A.}~\bibnamefont
  {Collinucci}}, \bibinfo {author} {\bibfnamefont {F.}~\bibnamefont {Denef}}, \
  and\ \bibinfo {author} {\bibfnamefont {M.}~\bibnamefont {Esole}},\ }\href
  {\doibase 10.1088/1126-6708/2009/02/005} {\bibfield  {journal} {\bibinfo
  {journal} {JHEP}\ }\textbf {\bibinfo {volume} {0902}},\ \bibinfo {pages}
  {005} (\bibinfo {year} {2009})},\ \Eprint {http://arxiv.org/abs/0805.1573}
  {arXiv:0805.1573 [hep-th]} \BibitemShut {NoStop}%
\bibitem [{\citenamefont {Blumenhagen}\ \emph {et~al.}(2009)\citenamefont
  {Blumenhagen}, \citenamefont {Braun}, \citenamefont {Grimm},\ and\
  \citenamefont {Weigand}}]{Blumenhagen:2008zz}%
  \BibitemOpen
  \bibfield  {author} {\bibinfo {author} {\bibfnamefont {R.}~\bibnamefont
  {Blumenhagen}}, \bibinfo {author} {\bibfnamefont {V.}~\bibnamefont {Braun}},
  \bibinfo {author} {\bibfnamefont {T.~W.}\ \bibnamefont {Grimm}}, \ and\
  \bibinfo {author} {\bibfnamefont {T.}~\bibnamefont {Weigand}},\ }\href
  {\doibase 10.1016/j.nuclphysb.2009.02.011} {\bibfield  {journal} {\bibinfo
  {journal} {Nucl.Phys.}\ }\textbf {\bibinfo {volume} {B815}},\ \bibinfo
  {pages} {1} (\bibinfo {year} {2009})},\ \Eprint
  {http://arxiv.org/abs/0811.2936} {arXiv:0811.2936 [hep-th]} \BibitemShut
  {NoStop}%
\bibitem [{\citenamefont {Arkani-Hamed}\ \emph {et~al.}(2005)\citenamefont
  {Arkani-Hamed}, \citenamefont {Dimopoulos},\ and\ \citenamefont
  {Kachru}}]{ArkaniHamed:2005yv}%
  \BibitemOpen
  \bibfield  {author} {\bibinfo {author} {\bibfnamefont {N.}~\bibnamefont
  {Arkani-Hamed}}, \bibinfo {author} {\bibfnamefont {S.}~\bibnamefont
  {Dimopoulos}}, \ and\ \bibinfo {author} {\bibfnamefont {S.}~\bibnamefont
  {Kachru}},\ }\href@noop {} {\  (\bibinfo {year} {2005})},\ \Eprint
  {http://arxiv.org/abs/hep-th/0501082} {arXiv:hep-th/0501082 [hep-th]}
  \BibitemShut {NoStop}%
\bibitem [{\citenamefont {Dine}\ and\ \citenamefont
  {Anisimov}(2005)}]{Dine:2004cq}%
  \BibitemOpen
  \bibfield  {author} {\bibinfo {author} {\bibfnamefont {M.}~\bibnamefont
  {Dine}}\ and\ \bibinfo {author} {\bibfnamefont {A.}~\bibnamefont
  {Anisimov}},\ }\href {\doibase 10.1088/1475-7516/2005/07/009} {\bibfield
  {journal} {\bibinfo  {journal} {JCAP}\ }\textbf {\bibinfo {volume} {0507}},\
  \bibinfo {pages} {009} (\bibinfo {year} {2005})},\ \Eprint
  {http://arxiv.org/abs/hep-ph/0405256} {arXiv:hep-ph/0405256 [hep-ph]}
  \BibitemShut {NoStop}%
\bibitem [{\citenamefont {Jeong}\ and\ \citenamefont
  {Takahashi}(2013)}]{Jeong:2013xta}%
  \BibitemOpen
  \bibfield  {author} {\bibinfo {author} {\bibfnamefont {K.~S.}\ \bibnamefont
  {Jeong}}\ and\ \bibinfo {author} {\bibfnamefont {F.}~\bibnamefont
  {Takahashi}},\ }\href {\doibase 10.1016/j.physletb.2013.10.061} {\bibfield
  {journal} {\bibinfo  {journal} {Phys.Lett.}\ }\textbf {\bibinfo {volume}
  {B727}},\ \bibinfo {pages} {448} (\bibinfo {year} {2013})},\ \Eprint
  {http://arxiv.org/abs/1304.8131} {arXiv:1304.8131 [hep-ph]} \BibitemShut
  {NoStop}%
\bibitem [{\citenamefont {Higaki}\ \emph
  {et~al.}(2014{\natexlab{b}})\citenamefont {Higaki}, \citenamefont {Jeong},\
  and\ \citenamefont {Takahashi}}]{Higaki:2014ooa}%
  \BibitemOpen
  \bibfield  {author} {\bibinfo {author} {\bibfnamefont {T.}~\bibnamefont
  {Higaki}}, \bibinfo {author} {\bibfnamefont {K.~S.}\ \bibnamefont {Jeong}}, \
  and\ \bibinfo {author} {\bibfnamefont {F.}~\bibnamefont {Takahashi}},\ }\href
  {\doibase 10.1016/j.physletb.2014.05.014} {\bibfield  {journal} {\bibinfo
  {journal} {Phys.Lett.}\ }\textbf {\bibinfo {volume} {B734}},\ \bibinfo
  {pages} {21} (\bibinfo {year} {2014}{\natexlab{b}})},\ \Eprint
  {http://arxiv.org/abs/1403.4186} {arXiv:1403.4186 [hep-ph]} \BibitemShut
  {NoStop}%
\bibitem [{\citenamefont {Kitajima}\ and\ \citenamefont
  {Takahashi}(2014)}]{Kitajima:2014xla}%
  \BibitemOpen
  \bibfield  {author} {\bibinfo {author} {\bibfnamefont {N.}~\bibnamefont
  {Kitajima}}\ and\ \bibinfo {author} {\bibfnamefont {F.}~\bibnamefont
  {Takahashi}},\ }\href@noop {} {\  (\bibinfo {year} {2014})},\ \Eprint
  {http://arxiv.org/abs/1411.2011} {arXiv:1411.2011 [hep-ph]} \BibitemShut
  {NoStop}%
\bibitem [{\citenamefont {Weinberg}(1989)}]{Weinberg:1988cp}%
  \BibitemOpen
  \bibfield  {author} {\bibinfo {author} {\bibfnamefont {S.}~\bibnamefont
  {Weinberg}},\ }\href {\doibase 10.1103/RevModPhys.61.1} {\bibfield  {journal}
  {\bibinfo  {journal} {Rev.Mod.Phys.}\ }\textbf {\bibinfo {volume} {61}},\
  \bibinfo {pages} {1} (\bibinfo {year} {1989})}\BibitemShut {NoStop}%
\bibitem [{\citenamefont {Banks}\ \emph {et~al.}(1991)\citenamefont {Banks},
  \citenamefont {Dine},\ and\ \citenamefont {Seiberg}}]{Banks:1991mb}%
  \BibitemOpen
  \bibfield  {author} {\bibinfo {author} {\bibfnamefont {T.}~\bibnamefont
  {Banks}}, \bibinfo {author} {\bibfnamefont {M.}~\bibnamefont {Dine}}, \ and\
  \bibinfo {author} {\bibfnamefont {N.}~\bibnamefont {Seiberg}},\ }\href
  {\doibase 10.1016/0370-2693(91)90561-4} {\bibfield  {journal} {\bibinfo
  {journal} {Phys.Lett.}\ }\textbf {\bibinfo {volume} {B273}},\ \bibinfo
  {pages} {105} (\bibinfo {year} {1991})},\ \Eprint
  {http://arxiv.org/abs/hep-th/9109040} {arXiv:hep-th/9109040 [hep-th]}
  \BibitemShut {NoStop}%
\bibitem [{\citenamefont {Kallosh}\ \emph {et~al.}(2014)\citenamefont
  {Kallosh}, \citenamefont {Linde},\ and\ \citenamefont
  {Vercnocke}}]{Kallosh:2014vja}%
  \BibitemOpen
  \bibfield  {author} {\bibinfo {author} {\bibfnamefont {R.}~\bibnamefont
  {Kallosh}}, \bibinfo {author} {\bibfnamefont {A.}~\bibnamefont {Linde}}, \
  and\ \bibinfo {author} {\bibfnamefont {B.}~\bibnamefont {Vercnocke}},\ }\href
  {\doibase 10.1103/PhysRevD.90.041303} {\bibfield  {journal} {\bibinfo
  {journal} {Phys.Rev.}\ }\textbf {\bibinfo {volume} {D90}},\ \bibinfo {pages}
  {041303} (\bibinfo {year} {2014})},\ \Eprint {http://arxiv.org/abs/1404.6244}
  {arXiv:1404.6244 [hep-th]} \BibitemShut {NoStop}%
\bibitem [{\citenamefont {Ade}\ \emph {et~al.}(2015)\citenamefont {Ade} \emph
  {et~al.}}]{Ade:2015tva}%
  \BibitemOpen
  \bibfield  {author} {\bibinfo {author} {\bibfnamefont {P.}~\bibnamefont
  {Ade}} \emph {et~al.} (\bibinfo {collaboration} {BICEP2, Planck}),\ }\href
  {\doibase 10.1103/PhysRevLett.114.101301} {\bibfield  {journal} {\bibinfo
  {journal} {Phys.Rev.Lett.}\ }\textbf {\bibinfo {volume} {114}},\ \bibinfo
  {pages} {101301} (\bibinfo {year} {2015})},\ \Eprint
  {http://arxiv.org/abs/1502.00612} {arXiv:1502.00612 [astro-ph.CO]}
  \BibitemShut {NoStop}%
\end{thebibliography}%
\end{document}